\documentclass{article}
\usepackage{amsmath} 
\usepackage{authblk} 
\usepackage{array}
\usepackage{graphicx} 
\usepackage[authoryear,round]{natbib}
\usepackage{url}
\usepackage{indentfirst}
\usepackage{fancyhdr}
\usepackage{blindtext} 
\usepackage{tabularx}
\usepackage{booktabs}
\usepackage{geometry}
\usepackage{longtable}
\usepackage{caption}
\usepackage{threeparttable} 
\usepackage{float}

\fancypagestyle{plain}{
    \fancyhf{} 
    \fancyhead[R]{\textit{LLM and Futures Price Factors in China} \quad \thepage} 
}

\title{Large Language Models and Futures Price Factors in China}
\author{}
\author{
Yuhan Cheng \thanks{School of Management, Shandong University, China. Email: chengyuhan@sdu.edu.cn} \ \
Yanchu Liu \thanks{Corresponding author. Lingnan College, Sun Yat-sen University, China.  Email: liuych26@mail.sysu.edu.cn} \ \
Heyang Zhou \thanks{Lingnan College, Sun Yat-sen University. Email: zhouhy97@mail2.sysu.edu.cn}}

\date{} 
\begin{document}

\maketitle



\begin{abstract}
We leverage the capacity of large language models such as Generative Pre-trained Transformer (GPT) in constructing factor models for Chinese futures markets. We successfully obtain 40 factors to design single-factor and multi-factor portfolios through long-short and long-only strategies, conducting backtests during the in-sample and out-of-sample period. Comprehensive empirical analysis reveals that GPT-generated factors deliver remarkable Sharpe ratios and annualized returns while maintaining acceptable maximum drawdowns. Notably, the GPT-based factor models also achieve significant alphas over the IPCA benchmark. Moreover, these factors demonstrate significant performance across extensive robustness tests, particularly excelling after the cutoff date of GPT's training data.
\end{abstract}


\textbf{Keywords:} Large language models, Chinese futures markets, factor analysis, zero-shot learning

\section{Introduction}
Risk factor models are fundamental tools in finance, designed to explain investment returns and evaluate risk. These models distill the complexities of capital markets into key factors that drive asset returns across various regions (Sharpe, 1964; Fama and French, 1992; Carhart, 1997; Fama and French, 2015; Fama and French, 2018; Hou et al., 2019). Over the years, technological advancements—particularly in machine learning and artificial intelligence—have transformed factor modeling by uncovering deep, non-linear patterns in vast datasets that were previously inaccessible (Kelly and Xiu, 2023). These breakthroughs have enhanced the agility and predictive accuracy of models. A cutting-edge development in finance is the application of large language models (LLMs), such as GPT series model, which are increasingly adept at analyzing unstructured data sources like news articles and social media. This capability enables more granular insights into market sentiment, contributing to the development of more powerful and predictive factor models. The adoption of LLMs underscores the growing convergence of finance and advanced computational techniques, continually reshaping investment strategies and risk management.

Despite these advancements, the Chinese futures market—a key representative of emerging markets—faces unique challenges in factor research. These challenges stem from two main facts: first, the potential differences in pricing rules and factor systems between the Chinese and U.S. markets, as validated by research in equity markets showing different factor relevances (Liu et al.,2019); and second, the limited availability of data in China, which makes it difficult for data-driven models like deep learning to perform effectively. Given these challenges, the use of GPT or, more broadly, LLMs offers a promising solution. The extensive knowledge base, reasoning capabilities, and code-generating abilities of these models can assist in overcoming the obstacles of factor research in the Chinese market. Specifically, LLMs offer two key advantages: (1) they can perform zero-shot learning, meaning they can generate insights without relying on extensive historical data; and (2) the time and cost associated with generating new ideas using GPT are significantly lower than those incurred by human researchers. 

This paper proposes applying LLM-based factor modeling as a solution not only for the Chinese market but also for any capital market where factor research is underdeveloped or particularly challenging. Our method begins by providing GPT-4 (hereinafter referred to as GPT) with specific data features, such as trading volume, open-close prices, and unique inputs like the basis and spot prices in the futures market. It is important to note that the GPT model used in our research was trained on data up to April 2023, which allows us to examine its genuine out-of-sample performance. To evaluate the efficacy of the GPT-generated factor models, we also employ the Instrumented Principal Component Analysis (IPCA) model as a benchmark, following the methodology of \cite{kelly2017}. This comparison aims to assess the performance of our GPT-derived factors against a well-established factor model in capturing the nuances of the Chinese futures market. 

In the broader field of asset pricing research, there is considerable concern about the potential for certain methods to introduce forward-looking bias. To address this issue in our work, we manually review and adjust the code generated by GPT to prevent such bias. Moreover, we do not view the future information inherently present in GPT itself as a significant concern. The key distinction in our approach lies in its departure from traditional GPT-based investment applications, which often rely on information streams—such as analyzing text, images, or data feeds—to predict futures market movements. Instead, we employ zero-shot reasoning, which operates without using any samples. This eliminates the risk of GPT cross-referencing user-provided data with its own training database, a common concern when directly utilizing data for stock predictions. Moreover, since GPT's training data was limited to information available up to April 2023 at the time of our research, we leverage this explicit cutoff characteristic to conduct a rigorous out-of-sample analysis. This means that GPT had absolutely no exposure to any data generated after this point in time, thereby allowing us to define the subsequent period as a strictly "out-of-sample" window for evaluation purposes. Specifically, we evaluate the factors' performance over an approximately one-year period following this cutoff date, during which GPT had no access to any relevant information. This approach effectively eliminates the risk of information leakage or data snooping, thereby mitigating the potential issue of forward-looking bias in factor construction. The results demonstrate consistently robust performance, aligning with our primary findings and underscoring the effectiveness of our methodology. This inherent advantage of LLMs, particularly their fixed training cutoff, represents a significant improvement over traditional factor models, where the risk of forward-looking bias is often more challenging to address.

Empirical analysis demonstrates that factors constructed by GPT are highly effective in building both long-short and long-only single-factor investment portfolios, with over half delivering positive annualized returns and Sharpe ratios alongside lower drawdowns, indicating strong risk-adjusted performance. Additionally, these factors exhibit strong alpha compared to the IPCA model, often surpassing significance levels, confirming their capacity for alpha generation and effective risk mitigation.Furthermore, multi-factor portfolios incorporating GPT-generated factors show similarly strong returns, and statistically significant alphas, underscoring their versatility and reliability in enhancing investment strategies. This success highlights GPT's power as a distinguished tool for factor construction in futures markets.

Our research enhances the application of LLMs within the financial sector by focusing on their innovative integration into the underexplored Chinese derivatives market. By deploying the advanced capabilities of GPT, this study not only fills a notable gap in the academic literature on asset pricing in this specific market but also pioneers the combination of LLMs with the IPCA model. Firstly, it greatly enriches the factor sets available for research in developing countries, where the study of factors in futures markets is notably limited. Secondly, it introduces a novel data-free methodology that is especially suitable for emerging markets with limited and short-term data, leveraging the knowledge and reasoning capabilities of LLMs. Thirdly, it refreshes our understanding of the pricing mechanisms within the Chinese futures market by providing not only valuable factors but also detailed explanations of the principles behind them. Lastly, it presents a new entry point for examining market efficiency in futures markets, contributing to the literature on market efficiency through the innovative integration of LLMs. These contributions extend the applications of LLMs beyond their traditional uses, demonstrating their potential to revolutionize factor analysis and asset pricing in financial markets.

The paper proceeds as follows. Section 2 introduces the literature review about the background of financial factor models, the application of machine learning in financial markets and the introduction of LLMs. Section 3 briefly describes the process of GPT's factor generation and Section 4 explains our methodology for performance evaluation. Section 5 and 6 summarize the factors and data and our empirical results, respectively. Section 7 summarizes the whole information that GPT shows, and we also make robustness tests in Seciton 8. Section 9 discusses different settings of prompt engineering and Section 10 concludes.

\section{Literature Review}
\subsection{Financial Factor Models}
The use of risk factors in financial markets traces back to the 1960s, originating from the groundbreaking contributions of the Capital Asset Pricing Model (CAPM) introduced by \cite{sharpe1964}, where risk was exclusively linked to the market factor. This concept was further developed by \cite{fama1992}, who added two more factors to the model, establishing the renowned Fama-French three-factor model, a foundational pillar in the field of quantitative finance. \cite{carhart1997} added momentum as a fourth factor to the Fama-French model, which captures the tendency of stocks that have performed well in the past to continue performing well in the near future. Continuing to the Fama-French model, \cite{FAMA2015} further refined their approach by introducing two additional factors, thus creating the Fama-French five-factor model. Meanwhile, \cite{hxz2015} proposed the q-factor model, which incorporates market, size, investment, and profitability factors. In a related effort to assess and rank asset pricing models, \cite{fama2018} explored the maximum squared Sharpe ratio as a metric for evaluating factor models, considering both nested and non-nested frameworks, including a six-factor extension of the Fama-French model that incorporates momentum. Building on these advancements, \cite{bfsj2018} compared multiple asset pricing models, demonstrating the importance of momentum, value, and profitability factors updated monthly. \cite{SY2016} introduced mispricing factors that aggregate anomalies to improve explanatory power, while \cite{dkd2019} proposed behavioral factors capturing investor psychology at short and long horizons, outperforming traditional models in explaining return anomalies. More recently, \cite{hxz2020} enhanced the q-factor model with an expected growth factor, improving its explanatory power and outperforming the Fama–French six-factor model. 

\subsection{Machine Learning in Financial Markets}
This rapidly developing approach in asset pricing heavily utilizes machine learning to analyze complex, high-dimensional data, significantly expanding beyond traditional revenue metrics. The method, as explored by \cite{gu2020}, employs sophisticated algorithms such as random forests and deep learning, addressing challenges like multicollinearity and enhancing the modeling of stock returns using diverse company characteristics. In a comprehensive review, \cite{kx2023} highlight key advancements and futures directions in machine learning applications within financial markets. Focusing on conditional factor models, \cite{lnm2017} and \cite{fjn2020} utilized sophisticated techniques like Partial Least Squares (PLS) and LASSO to address statistical complexities involving numerous predictors. \cite{ksns2018} demonstrated that a select group of principal components from 15 unique portfolios could effectively assess these portfolios' values, though resulting in negligible alpha. Moreover, \cite{MURRAY2024} applied machine learning to validate the efficient market hypothesis by predicting stock returns based on historical data, which accurately forecasted future stock returns across a variety of markets, yet the model showed limitations with certain asset types. The recent introduction of a novel Feature Set Data-Driven Search framework (FETCH) by \cite{li2023learning} and a novel automated feature generation method (OpenFE) by \cite{zhang2023} further exemplifies the ongoing innovation in machine learning within the financial sector. These tools extend the capabilities of machine learning systems in feature generation and selection, achieving performance on par with expert-level financial analysts. Additionally, the reinforcement learning-based model proposed by \cite{Yu_2023} represents a significant shift towards more adaptive and performance-optimized factor creation strategies, moving away from traditional methods to a more dynamic and tailored approach.

\subsection{Large Language Models}
LLMs like GPT represent a significant advancement in the field of artificial intelligence, particularly within the realm of natural language processing. These models are trained on extensive datasets from a diverse range of sources, allowing them to generate coherent, contextually relevant text based on the prompts they receive. LLMs function by predicting the next word in a sequence, given all the previous words, thus simulating a form of human-like understanding of language.

The application of large language models (LLMs) in the financial sector represents a transformative advancement in leveraging artificial intelligence for complex domains. Despite their proficiency in natural language processing, LLMs face challenges in finance due to the domain's intricate dynamics \cite{zhxy2023}. Domain-specific training has proven effective, as demonstrated by BloombergGPT, which leverages extensive financial data to enhance performance in financial contexts \cite{wild2023}, and by frameworks tailored for the Chinese financial market, addressing unique regional characteristics \cite{jyg2023}. LLMs have also shown versatility in specialized financial tasks, including economic forecasting and complex reasoning for financial analytics \cite{lwzzw2023, sghm2023}. Notably, GPT-based models have been utilized for innovative applications, such as deriving firm-level investment scores from corporate communications to predict capital expenditures \cite{NBERw2024}, generating predictive financial factors from financial statement structures without relying on numerical data \cite{cheng2024}, and predicting equity risk premiums in the Chinese market, outperforming traditional methods like the bag-of-words approach \cite{MA2024}. Furthermore, GPT enhances financial sentiment analysis, strategic decision-making, and market trend decoding \cite{gjk2023, gyxy2023}, while also supporting forecasting of excess returns in commodity futures through text-based indicators \cite{gwwz2025}. Preprocessing techniques and multitask prompting significantly boost LLM performance in financial tasks \cite{lwdc2023}, and GPT’s role in financial research, banking, and customer interactions underscores its potential to revolutionize advisory services \cite{dmlb2023}. However, ethical considerations and future research directions remain critical, as highlighted by \cite{klmy2023}, emphasizing the need for responsible deployment of LLMs in finance.

Our work is most closely related to \cite{yckt2024}. This seminal work explores the integration of the GPT computational model with innovative autonomous factor generation techniques, leading to the creation of high-return investment factors within the equity market. Additionally, \cite{aomh2024} investigate the effectiveness of GPT in providing financial advice compared to robo-advisors, concluding that GPT delivers superior recommendations aligned with investor profiles and benchmarks.  \cite{mpjv2024} examine GPT's potential in stock picking, highlighting its positive impact on investment strategy and returns.


Our study diverges from \cite{Chen2023} by focusing on the application of GPT-generated factors to build a factor model, while \cite{Chen2023} investigates GPT's ability to identify macroeconomic news content that influences the aggregate stock market, our research leverages the advanced capabilities of the GPT algorithm to generate financial factors specifically tailored to the Chinese futures market. This distinctive approach underscores the unique strengths of GPT in formulating high-return investment strategies, demonstrating the model's versatility beyond merely analyzing news content and macroeconomic correlations. By focusing on the generation and performance of these factors within the futures market, our study highlights GPT's practical applicability and robustness in a different financial domain. This not only broadens the understanding of GPT's potential but also showcases its effectiveness in diverse financial settings, providing valuable insights for future applications.

Building upon the foundation laid, our research extends and innovates on their methodologies by applying newer versions of language models and expanding the scope of factor generation to include additional markets and a broader array of financial instruments. We aim to explore the scalability of the GPT-based factor generation framework and its applicability in varying market conditions, thus providing a more comprehensive understanding of its potential and limitations in global financial contexts. 

\subsection{Factors in Futures Markets}
In the realm of commodity futures markets, factor models have been instrumental in understanding and predicting price behaviors. \cite{JR1985} extended the ICAPM to test futures prices for corn, wheat, and soybeans, providing valuable parameter estimates despite rejecting the model. \cite{cgnl2006} employed an N-factor Gaussian model using all available price information, demonstrating strong performance across oil and copper futures. Initial models, such as those proposed by \cite{CORTAZAR2003215}, have been extended to incorporate features like convenience yields and spot price volatility, tailoring models to specific commodities, as seen in studies like \cite{ksat2012} on oil. \cite{wjnm2015} supported the Efficient Futures Market Hypothesis through a factor analytical approach, enhancing test power with panel data. Chong et al. (2017) identified liquidity as a critical pricing factor in commodity futures, while \cite{HD2019} highlighted the dynamic relationship between oil futures and economic activity.

In recent years, researchers have explored various extensions and applications of these models. \cite{kjkky2017} investigated the predictability of commodity futures momentum, while \cite{hykl2022} identified a trend factor that outperformed traditional momentum factors. Recent work has compared two-factor and three-factor models, refining them to better reflect market realities and the interplay of market forces, especially in multi-commodity scenarios, as highlighted by studies such as \cite{HH2000}. Challenges persist in fully capturing the complexities of market forces, particularly in multi-commodity settings, as noted by \cite{cgnl2006}. Complementing these efforts, \cite{dckas2013} explored whether common factors exist in individual commodity futures returns, utilizing cross-sectional models motivated by commodity pricing theories. In the Chinese context, \cite{zjz2020} conducted a broad analysis of commodity futures returns, identifying significant carry and momentum factors, and highlighted behavioral influences like herding effects. Similarly, \cite{jhx2021} demonstrated a monotonic, asymmetric negative relationship between skewness and expected returns, underscoring skewness as a distinct risk factor in China's commodity futures market. Furthermore, \cite{qpyk2020} showed that incorporating higher-order moments, particularly the systematic fourth-order moment, into portfolio strategies enhances performance. Extending beyond observed factor models, \cite{liu2025} proposed latent factor models for Chinese commodity futures markets using Instrumented Principal Component Analysis and Conditional Autoencoder methods, uncovering hidden drivers of futures prices with robust out-of-sample predictive accuracy. These studies collectively advance our understanding of pricing dynamics, hedging strategies, and economic implications in futures markets.

\section{Methodology}
This section outlines our methodology for generating and evaluating investment factors derived from GPT. We first describe the process of auto-generating novel factors using GPT and then assess their financial performance through a series of backtests. To evaluate the practical utility of these factors, we construct portfolios based on each generated factor and measure key performance indicators including annualized return, Sharpe ratio, and maximum drawdown. These metrics serve as the foundation for assessing risk-adjusted profitability. For benchmarking, we also compare the alpha generation ability of these factors against the IPCA model proposed by Kelly (2017), thereby situating our findings within a broader empirical finance context.

\subsection{GPT’s Auto-factor Generation}
We utilize GPT as an artificial financial analyst to generate novel, multi-dimensional financial factors to predict future stock returns, balancing generalizability and computational efficiency. To avoid misinterpretations from GPT's text-based outputs or mathematical notations, we restrict its access to our data architecture framework, excluding substantive data, and extract the computational algorithm directly from the model \cite{yckt2024}. GPT is tasked with defining a computational framework based on its understanding of dataset column labels and their meanings, comprising sequential cross-sectional data. It proposes a complex function, \textit{f}, to dynamically model the factor magnitude vector across all futures at time \textit{t}, leveraging market nuances at and before \textit{t}. A thorough manual review ensures the factor’s uniqueness and validity, preventing duplication of existing models and advancing financial research with innovative AI-driven analytical tools.

Our methodological approach involves guiding GPT step-by-step to systematically develop the computational blueprint for an innovative financial factor. This structured process begins with careful prompt engineering, explicitly instructing GPT to assume the role of a financial expert, thus aligning its analytical perspective closely with the intricacies of financial analysis. Next, we comprehensively brief GPT on the specific task of factor generation, clearly communicating the nature of the problem, the desired outcome, and the underlying data schema, while deliberately avoiding exposure to actual data to prevent potential biases. Based on this briefing, GPT generates the algorithmic representation of the factor in Python code, capitalizing on its expansive training to hypothesize a functional model $f$ that effectively captures complex financial relationships. Subsequently, we rigorously perform a forward-looking bias check, meticulously verifying that the factor does not inadvertently incorporate future data, ensuring its validity and applicability for real-time predictive modeling. To uphold our objective of innovation, GPT is explicitly directed to propose multifaceted factors distinct from those previously documented in financial literature. Lastly, we conduct a comprehensive manual review of the factors generated by GPT to confirm their novelty and academic contribution. This entire methodological workflow—from initial prompt engineering and factor generation to bias validation and originality assessment—is systematically documented and visually presented as a flowchart in Figure 1, clearly illustrating each step’s logic and the rationale guiding the AI-driven factor development process.

\begin{figure}[H]
  \centering  
  \includegraphics[width=0.8\textwidth]{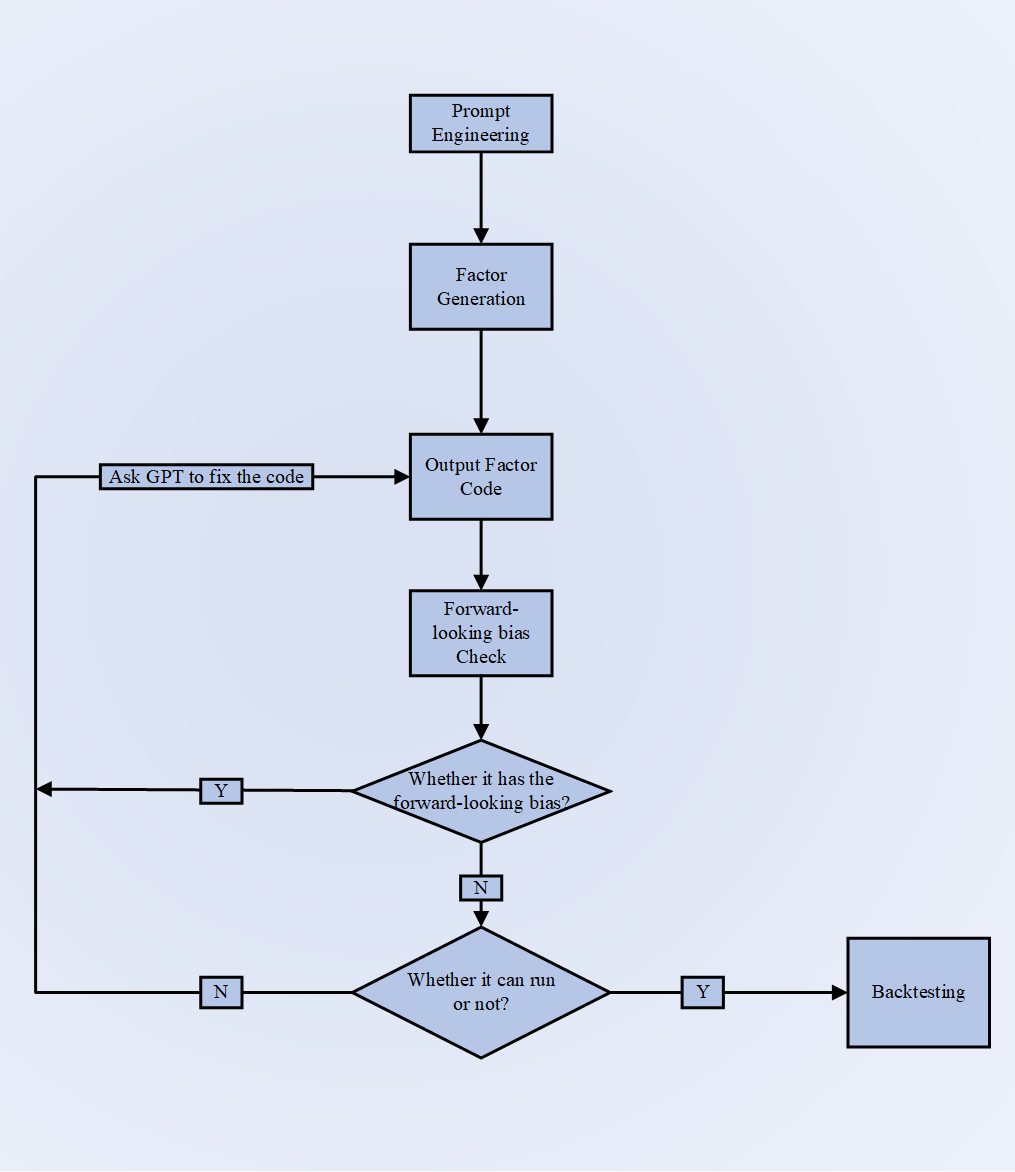}  
  \caption{Guiding GPT in factor generation. This figure illustrates how we direct GPT to produce useful codes for constructing factors and implement them in a way without worries about the forward-looking bias issue.}  
  \label{fig.image1}  
\end{figure}

In our study, the foundational step of our methodology is the “Prompt Engineering” phase, which plays a critical role in setting the direction and effectiveness of the factor generation process. This initial stage involves configuring GPT to assume the role of a financial expert. The precision of our prompt engineering is crucial because it defines the scope and capabilities of the AI model in the context of financial analysis, thereby influencing the subsequent steps of factor generation.Effective prompt engineering results in the generation of a more refined and contextually aware GPT, which then proceeds to the next stage of factor generation with a clear understanding of its task. The prompt we used was:

“You are a fund manager to invest in futures, and I need you to brainstorm new risk-factors for me. You have panel data for many futures stored as a data frame named “data”. I require you to provide me with a Python program that can generate a new column as a factor. Your factor should be original, novel and innovative. Think creatively. The data columns include futuresname, date, basis, spot price, futures premium and discount,open, high, low, close, volume, and amount, where you should use as least two or more. Also, You need to name the factor and explain how you construct it. However, please ensuring originality is the most important.” In this context, basis is the difference between the spot price and futures price of an asset, spot price is the current price at which an asset can be bought or sold for immediate delivery and futures premium and discount refer to whether the futures price is higher (premium) or lower (discount) than the spot price, reflecting market expectations.

As our research progressed, we prompted GPT with a simple request: “Please generate another factor.” Initially effective, GPT began introducing variables outside our data schema after about ten factors. To correct this, we refined the prompt to: “Please generate another novel and original factor and remember the data columns include futuresname, date, basis, spot price, futures premium and discount, open, high, low, close, volume, and amount, where you should use at least two or more,” successfully guiding GPT to generate relevant, schema-compliant factors. However, we also discovered that GPT-generated code often contained forward-looking bias, using future data inadvertently. To ensure model integrity, we implemented a forward-looking bias check before execution. Additionally, during operational testing, we encountered technical issues such as division by zero, infinite loops, and memory overflow causing segmentation faults. These required a rigorous validation process to identify and exclude factors with sparse or aberrant outputs, ensuring only robust, computationally sound factors were retained, thereby enhancing the reliability of our financial models.

\subsection{Construction of Different Factor Portfolios}
In constructing and evaluating factor-based portfolios, we begin with a single-factor model employing two investment strategies: a long-short and a long-only approach. Futures are ranked daily by their factor values; the long-short strategy takes long positions in the top 10\% and short positions in the bottom 10\%, while the long-only strategy focuses solely on the top decile. Building upon this, we extend to multi-factor portfolio construction using both static and dynamic strategies as outlined in \cite{yckt2024}. The static approach determines the polarity of each factor—positive or negative—based on its historical long-short returns. These polarities guide the aggregation of standardized factor signals across multiple dimensions, yielding a composite score for each asset. This methodology enables a flexible yet principled integration of multiple predictive signals in a unified portfolio framework.

To validate the robustness and effectiveness of factors, we divide the time series data of each factor into two segments. The first segment is used to construct the long-short portfolio, and its performance, particularly the annualized return rate, is employed to determine the sign (positive or negative) of the factor. This sign assignment is critical as it indicates whether the factor predicts futures performance in alignment with market movements. The second segment of the data is then used to build the actual investment portfolios to be evaluated, which focuses on generating performance metrics. By separating the dataset into two parts for different purposes—model building and performance evaluation—we ensure that our analysis is robust and that the factor's effectiveness in real-world scenarios is accurately assessed. This methodology not only highlights the predictability of the generated factors but also establishes a systematic approach to testing their practical applicability in real market.


On the other hand, the dynamic method involves daily adjustments to the polarity of factor values based on cumulative historical returns. This approach requires recalculating the daily factor values with the historical return-based sign, incorporating real-time data to refine the portfolio continuously. Subsequently, the average factor value for each futures is computed using the updated values. This dynamic strategy allows for a more responsive portfolio that adjusts to market conditions in real-time, potentially enhancing returns and mitigating risks associated with market volatility. In the multi-factor portfolio investment model, both the long-only strategy and the long-short one are also implemented. 

In addition, we evaluate the alpha of our constructed portfolios using the IPCA model as a benchmark, specifically adopting the IPCA-5 variant due to its superior empirical performance. In implementing the IPCA benchmark, we ensure consistency by aligning the raw input features of IPCA with those accessible to GPT. Furthermore, the training period for the IPCA model precisely matches the polarity adjustment period used in the GPT-generated factors, while the IPCA testing period aligns exactly with the backtesting interval of portfolios constructed from GPT-generated factors. Through comprehensive empirical analyses of IPCA models with varying numbers of factors (IPCA 1–5) within the Chinese futures market, we identify the IPCA-5 model as the best-performing benchmark. This rigorous methodological alignment allows for an accurate and detailed assessment of the effectiveness of our GPT-based investment strategies in delivering superior risk-adjusted returns.



\section{Factors and Data}
In this section, we delve into the intricacies of the factors generated by GPT and the datasets utilized in our study. By elucidating the underlying mechanisms and the breadth of data employed, we aim to provide a comprehensive understanding of the parameters that influence GPT's outputs. This analysis will lay the groundwork for a nuanced exploration of the model's performance and capabilities.

Harnessing the advanced computational capabilities of GPT, we generated a diverse array of factors, ultimately identifying 40 that seamlessly fit into our coding architecture. Each factor underwent a thorough evaluation process to verify its compatibility and relevance within our analytical framework, which ensured that the factors were robust and applicable across a range of market scenarios.


For the following empirical analysis, we collected daily market data for 104 leading and secondary continuous contracts, including all stock index futures and liquid commodity futures in China, from the Wind financial database. The data included variables such as basis, spot, futures premium and discount, open, high, low, close, volume, amount, futures name, and return, covering the period from 2010/01/04 to 2023/04/07. However, it should be noted that some futures data do not span the entire period above. Initially, we used data up to 2017/12/29, to generate trading signals. The backtesting period extended from 2018/01/02 to 2023/04/07, to evaluate the performance of factors. In this way we could ensure a robust assessment of the models' predictive capabilities under various market conditions. Elementary information for all futures contracts is detailed in Table A1, Online Appendix A.

Table \ref{ICIR} provides the IC (Information Coefficient) Mean and IR (Information Ratio) for all factors. IC is a statistical measure that indicates the correlation between the predicted and actual returns. A higher IC value suggests that the factor has a stronger predictive power for returns. An IC value close to zero indicates no predictive power, while a negative IC implies that the factor tends to make incorrect predictions. IR, on the other hand, is used to evaluate the efficiency of a factor, which is calculated as the ratio of the factor’s excess return to its volatility. The higher the IR, the more efficiently the factor generates returns relative to the risk it incurs. The outcome reveals that most factors demonstrate positive predictive power, with several showing exceptional performance. For instance, Factor 22 has a high IC of 0.1673 and an IR of 0.6756, indicating remarkable risk-return balance. Factor 34 is even more robust, with an IC of 0.3021 and an IR of 1.2645. In contrast, Factor 7 performs poorly, with a negative IC of -0.5821 and an IR of -3.5624, reflecting weak predictive power and high volatility. Overall, the analysis highlights that many factors offer reliable predictability, with some performing exceptionally well, which suggests that the GPT factors perform strongly based on IC and IR criteria.

\begin{table}[htbp]
\centering
\begin{threeparttable}
\caption{IC Mean and IR of Factors}
\label{ICIR}
\vspace{1em} 
\centering
\begin{tabular}{l@{\hskip 40pt}c@{\hskip 40pt}c@{\hskip 40pt}l@{\hskip 40pt}c@{\hskip 40pt}c}
\hline
\textbf{Factor} & \textbf{IC Mean} & \textbf{IR} & \textbf{Factor} & \textbf{IC Mean} & \textbf{IR} \\
\hline
Factor 1  & 0.0962  & 0.4008  & Factor 21 & 0.0022  & 0.0124 \\
Factor 2  & 0.0270  & 0.1344  & Factor 22 & 0.1673  & 0.6756 \\
Factor 3  & 0.0098  & 0.0378  & Factor 23 & 0.0114  & 0.0463 \\
Factor 4  & 0.0312  & 0.1467  & Factor 24 & -0.0096 & -0.0536 \\
Factor 5  & 0.0012  & 0.0053  & Factor 25 & 0.0008  & 0.0037 \\
Factor 6  & -0.0007 & -0.0044 & Factor 26 & 0.0134  & 0.0666 \\
Factor 7  & -0.5821 & -3.5624 & Factor 27 & 0.0254  & 0.1135 \\
Factor 8  & 0.0225  & 0.0756  & Factor 28 & 0.0428  & 0.2008 \\
Factor 9  & 0.1326  & 0.6152  & Factor 29 & -0.0081 & -0.0543 \\
Factor 10 & -0.0133 & -0.0575 & Factor 30 & 0.0009  & 0.0046 \\
Factor 11 & -0.0109 & -0.0466 & Factor 31 & 0.0876  & 0.4477 \\
Factor 12 & -0.0134 & -0.0578 & Factor 32 & -0.0075 & -0.0287 \\
Factor 13 & 0.0267  & 0.0971  & Factor 33 & 0.0169  & 0.0999 \\
Factor 14 & 0.0375  & 0.1881  & Factor 34 & 0.3021  & 1.2645 \\
Factor 15 & 0.0138  & 0.0741  & Factor 35 & 0.0106  & 0.0403 \\
Factor 16 & 0.0177  & 0.0879  & Factor 36 & -0.0046 & -0.0235 \\
Factor 17 & 0.0548  & 0.2829  & Factor 37 & -0.0264 & -0.1674 \\
Factor 18 & 0.0674  & 0.2941  & Factor 38 & 0.0735  & 0.3575 \\
Factor 19 & 0.0078  & 0.0408  & Factor 39 & 0.0158  & 0.0795 \\
Factor 20 & 0.0232  & 0.1074  & Factor 40 & -0.0007 & -0.0038 \\
\hline
\end{tabular}
\begin{tablenotes}[flushleft]\footnotesize
\item Notes: This table shows the Spearman IC mean and IR of factors created by GPT. Data covers the period from 2010/01/04 to 2023/04/07. A higher IC indicates that the factor more accurately predicts futures performance. The IC mean reflects the average predictive ability of the factor over time. A higher IR suggests that the factor not only has strong predictive power but also exhibits stability across different market conditions, making it a more reliable tool for portfolio construction and investment strategies.
\end{tablenotes}
\end{threeparttable}
\end{table}

\section{Empirical Results}
\subsection{In-Sample Performance}
In this section, we present the empirical analysis of the performance of various factors, specifically focusing on the single factor portfolio with long-short and long-only strategies. Key metrics for performance evaluation are the Annualized Return, Sharpe Ratio, and Max Drawdown. Table \ref{ISLS} exhibits empirical results for each factor of the long-short strategy.

The Annualized Return metric, representing the compounded yearly growth rate, provides insights into the profitability of our factors. Under the long-short framework, 28 factors exhibited positive returns, while 12 displayed negative returns. Factor 22 delivered exceptional performance with an annualized return of 7.4961 and the highest Sharpe ratio of 8.8570, significantly outperforming other factors, followed by Factor 9 with a return of 3.4037. Factors 31 and 28 also achieved strong returns exceeding 2, and overall, seven factors had annualized returns greater than 1. Analysis of the Sharpe ratio indicates a generally positive correlation between returns and risk-adjusted performance, despite notable exceptions such as Factor 7. Regarding downside risk, 31 factors maintained maximum drawdowns below 0.5; however, some experienced extreme drawdowns of up to 1.0. Among positively performing factors, 11 had drawdown rates exceeding 0.5, whereas those with high Sharpe ratios showed maximum drawdowns capped at 0.2843. Factor 17 emerged as particularly resilient, with the lowest drawdown of 0.0298, thus strengthening its risk-adjusted performance profile.

\begin{table}[htbp]
\centering
\begin{threeparttable}
\caption{In-Sample Performance of Factors: Long-Short Portfolio}
\label{ISLS}
\vspace{1em} 
\centering
\begin{tabularx}{\textwidth}{>{\raggedright\arraybackslash}p{2cm} >{\raggedright\arraybackslash}p{2cm} >{\raggedright\arraybackslash}p{2cm} >{\raggedright\arraybackslash}p{2cm} >{\raggedright\arraybackslash}p{2cm}>{\raggedright\arraybackslash}p{2cm}>{\raggedright\arraybackslash}p{2.5cm}}
\hline
\textbf{Factor} & \textbf{Annualized Return} & \textbf{Sharpe Ratio} & \textbf{Max Drawdown} & \textbf{Annualized Return*} & \textbf{Sharpe Ratio*} & \textbf{Max Drawdown*} \\
\hline
Factor 1  & 2.8014  & 5.9213  & 0.1311  & 2.3534  & 5.3749  & 0.1337  \\
Factor 2  & 0.2983  & 1.5785  & 0.2310  & 0.1447  & 0.8593  & 0.3241  \\
Factor 3  & 0.2569  & 1.1472  & 0.3745  & 0.1081  & 0.5758  & 0.4893  \\
Factor 4  & 0.4940  & 2.5008  & 0.1576  & 0.3173  & 1.7424  & 0.1797  \\
Factor 5  & 0.0125  & 0.1576  & 0.3191  & -0.1074  & -0.6849  & 0.4875  \\
Factor 6  & 0.0262  & 0.2416  & 0.3475  & -0.0953  & -0.5472  & 0.4955  \\
Factor 7  & -0.9983  & -27.3291  & 1.0000  & -0.9985  & -27.8760  & 1.0000  \\
Factor 8  & 0.0076  & 0.1701  & 0.6110  & -0.1117  & -0.2688  & 0.6857  \\
Factor 9  & 3.4037  & 7.9222  & 0.1138  & 2.8851  & 7.2588  & 0.1253  \\
Factor 10  & -0.1626  & -0.8166  & 0.6279  & -0.2618  & -1.4659  & 0.7884  \\
Factor 11  & -0.2167  & -0.9678  & 0.7411  & -0.3096  & -1.5257  & 0.8558  \\
Factor 12  & -0.0186  & 0.0154  & 0.4628  & -0.1348  & -0.5846  & 0.6055  \\
Factor 13  & 0.6720  & 1.9784  & 0.5157  & 0.4744  & 1.5281  & 0.5679  \\
Factor 14  & 0.5559  & 2.8752  & 0.1701  & 0.3719  & 2.0788  & 0.1914  \\
Factor 15  & 0.0674  & 0.4676  & 0.3618  & -0.0590  & -0.2713  & 0.4731  \\
Factor 16  & 0.3443  & 1.8264  & 0.1522  & 0.1853  & 1.0855  & 0.2463  \\
Factor 17  & 1.9102  & 8.2322  & 0.0298  & 1.5669  & 7.2709  & 0.0362  \\
Factor 18  & 1.1311  & 4.6638  & 0.0717  & 0.8794  & 3.9019  & 0.0931  \\
Factor 19  & 0.0797  & 0.5289  & 0.3206  & -0.0481  & -0.1969  & 0.4084  \\
Factor 20  & -0.2786  & -2.1723  & 0.8189  & -0.3641  & -3.0393  & 0.9036  \\
Factor 21  & 0.0859  & 0.5820  & 0.3163  & -0.0427  & -0.1826  & 0.4491  \\
Factor 22  & 7.4961  & 8.8570  & 0.1444  & 6.4980  & 8.3447  & 0.1528  \\
Factor 23  & 0.0223  & 0.2146  & 0.4095  & -0.0987  & -0.5172  & 0.5567  \\
Factor 24  & 0.0214  & 0.2210  & 0.1842  & -0.0996  & -0.6807  & 0.4405  \\
Factor 25  & 0.1084  & 0.6687  & 0.2089  & -0.0228  & -0.0418  & 0.3840  \\
Factor 26  & -0.0112  & 0.0258  & 0.3469  & -0.1283  & -0.6821  & 0.5525  \\
Factor 27  & -0.2919  & -1.3334  & 0.9514  & -0.3759  & -1.8636  & 0.9701  \\
Factor 28  & 2.2622  & 4.9251  & 0.2843  & 1.8776  & 4.4145  & 0.2961  \\
Factor 29  & -0.1288  & -0.8101  & 0.6476  & -0.2320  & -1.6213  & 0.7679  \\
Factor 30  & 0.0341  & 0.2971  & 0.2348  & -0.0883  & -0.5334  & 0.3857  \\
Factor 31  & 2.5680  & 6.6688  & 0.2194  & 2.1475  & 6.0194  & 0.2265  \\
Factor 32  & -0.1028  & -0.4217  & 0.4709  & -0.2091  & -1.0316  & 0.7138  \\
Factor 33  & 0.3014  & 1.7884  & 0.1636  & 0.1475  & 0.9700  & 0.1835  \\
Factor 34  & -0.9688  & -18.0299  & 1.0000  & -0.9726  & -18.6931  & 1.0000  \\
Factor 35  & 0.1832  & 1.0268  & 0.4187  & 0.0432  & 0.3256  & 0.4864  \\
Factor 36  & -0.0950  & -0.4828  & 0.4898  & -0.2022  & -1.2035  & 0.7101  \\
Factor 37  & -0.1863  & -1.5706  & 0.6516  & -0.2828  & -2.5692  & 0.8159  \\
Factor 38  & 1.2041  & 3.8101  & 0.0935  & 0.9439  & 3.2205  & 0.0983  \\
Factor 39  & 0.1495  & 0.8739  & 0.2615  & 0.0134  & 0.1636  & 0.4114  \\
Factor 40  & -0.0141  & 0.0036  & 0.2397  & -0.1309  & -0.7285  & 0.5609  \\
\hline
\end{tabularx}
\begin{tablenotes}[flushleft]\footnotesize
\item Notes: This table shows In-Sample performance of factors in long-short portfolios, calculated by GPT. In-Sample means the backtesting period occurs before the cutoff date of GPT’s training data. It involves long positions in the top 10\% of futures by factor values and short positions in the bottom 10\%. Data covers the period from 2018/01/02 to 2023/04/07, with historical data from 2010 to 2021 used for reference. Symbol \* indicates results adjusted for transaction fees.
\end{tablenotes}
\end{threeparttable}
\end{table}

The long-only strategy, detailed in Table \ref{ISLO}, provides a comprehensive analysis of annual returns, Sharpe ratios, and maximum drawdowns for each factor. Most factors generate positive returns, with Factor 18 notably achieving substantial growth at 5.7249, followed by Factor 22 at 3.7633, and Factors 31 and 9 with returns of 1.1063 and 1.1694, respectively. Overall, five factors surpass a 1.0 annual return threshold, demonstrating strong performance. Factor 18 also excels in risk-adjusted returns, posting an impressive Sharpe ratio of 11.4682, while Factors 22 and 17 achieve Sharpe ratios above 5.0. In total, 19 factors deliver Sharpe ratios exceeding 1.0, signaling favorable risk-adjusted returns across most of the portfolio. However, Factors 7 and 34 significantly underperform, both exhibiting negative returns and maximum drawdowns of 1.0. Comparing long-only to long-short strategies, the long-only portfolio generally experiences higher drawdowns, with Factor 34 being particularly affected. Nevertheless, several factors demonstrate resilience, notably Factor 17 with a minimal drawdown of 0.08, and Factors 18 and 38 exhibiting drawdowns of 0.0639 and 0.1403, respectively, reflecting stability amidst market volatility.

After adjusting for transaction fees (i.e., 0.025\%), we observe that the majority of factors show only minor changes in their performance metrics, further emphasizing their robustness in real-world applications. Using the results of the long-short portfolio as an example, Factor 1's annualized return drops from 2.8014 to 2.3534 after adjustment, while its Sharpe ratio remains comparably high, indicating minimal impact on the factor’s risk-adjusted returns. However, certain factors experience more pronounced differences. For example, Factor 5 sees a shift from a positive annualized return of 0.0125 to a negative return of -0.1074 after transaction fee adjustments. Overall, while transaction fees do affect the returns of some factors more than others, the majority of factors remain stable, demonstrating their potential for application in real-world trading strategies despite the presence of costs.

Table \ref{ISA} reports alpha returns from the IPCA benchmark under long-short and long-only strategies. In the long-short approach, Factors 1, 2, 4, 9, 13, 14, 17, 18, 22, 28, 31, and 38 show 1\% significance, with Factor 9 (0.2551***) and Factor 28 (0.2281***) exhibiting particularly high alphas; Factor 9 also achieves a strong long-only alpha of 0.6112***. Factors 22 and 31 further confirm this trend with alphas of 0.3812*** and 0.2316***, respectively. Factors 3, 39, and 20 are significant at 5\%, and Factors 10 and 27 at 10\%. Notably, Factor 7 shows a large negative alpha (-1.1335***). Under the long-only strategy, many factors remain significant at 1\%, with Factor 18 (1.3927***) and Factor 22 (1.1835***) leading, while Factors 35, 7, and 37 are significant at 5\%, and Factors 21, 23, and 27 at 10\%. The consistent significance of Factors 9, 18, 22, and 28 across both strategies highlights their robustness and strong outperformance relative to the IPCA benchmark, reinforcing their value in building resilient, high-performing factor-based portfolios.

Following the articulated procedure for multi-factor portfolio configuration and its subsequent daily recalibration, the return dynamics are encapsulated in Table \ref{ISM}. For the static long-only portfolio, it achieved an annual return of 0.6528, paired with a Sharpe ratio of 2.26. The maximum drawdown, recorded at 0.2704, aligns with most factors evaluated under long-only strategies, indicating a moderate level of risk. In contrast, the static long-short portfolio experienced an annual return of 0.8111, which is notably higher than the long-only strategy. Furthermore, the Sharpe ratio also improved to 3.0218, and the maximum drawdown decreased significantly to 0.1623. These figures suggest that the risk was effectively mitigated, offering a more stable investment profile with the long-short strategy.

\begin{table}[H]
\centering
\begin{threeparttable}
\caption{In-Sample Performance of Factors: Long-Only Portfolio}
\label{ISLO}
\vspace{1em} 
\begin{tabularx}{\textwidth}{>{\raggedright\arraybackslash}p{2cm} >{\raggedright\arraybackslash}p{2cm} >{\raggedright\arraybackslash}p{2cm} >{\raggedright\arraybackslash}p{2cm} >{\raggedright\arraybackslash}p{2cm}>{\raggedright\arraybackslash}p{2cm}>{\raggedright\arraybackslash}p{2.5cm}}
\hline
\textbf{Factor} & \textbf{Annualized Return} & \textbf{Sharpe Ratio} & \textbf{Max Drawdown} & \textbf{Annualized Return*} & \textbf{Sharpe Ratio*} & \textbf{Max Drawdown*} \\
\hline
Factor 1 & 1.0547 & 3.1265 & 0.1763 & 0.8120 & 2.6011 & 0.1805 \\
Factor 2 & 0.2069 & 1.0478 & 0.3472 & 0.0641 & 0.4125 & 0.4517 \\
Factor 3 & 0.1939 & 0.8493 & 0.3855 & 0.0526 & 0.3322 & 0.4942 \\
Factor 4 & 0.3261 & 1.4766 & 0.2235 & 0.1692 & 0.8636 & 0.2681 \\
Factor 5 & 0.0073 & 0.1278 & 0.3256 & -0.1120 & -0.6110 & 0.4880 \\
Factor 6 & 0.0585 & 0.4072 & 0.3246 & -0.0668 & -0.2958 & 0.4374 \\
Factor 7 & -0.9579 & -14.5503 & 1.0000 & -0.9630 & -15.1369 & 1.0000 \\
Factor 8 & 0.0457 & 0.2993 & 0.6196 & -0.0781 & -0.1070 & 0.6815 \\
Factor 9 & 1.1694 & 3.7280 & 0.2179 & 0.9133 & 3.1398 & 0.2410 \\
Factor 10 & -0.0456 & -0.2129 & 0.3890 & -0.1586 & -1.0037 & 0.5867 \\
Factor 11 & 0.0226 & 0.2169 & 0.5052 & -0.0985 & -0.2585 & 0.5805 \\
Factor 12 & 0.0467 & 0.3434 & 0.2553 & -0.0772 & -0.3555 & 0.3992 \\
Factor 13 & 0.5510 & 1.4176 & 0.5909 & 0.3676 & 1.0618 & 0.6311 \\
Factor 14 & 0.3828 & 1.8260 & 0.2466 & 0.2192 & 1.1529 & 0.2816 \\
Factor 15 & 0.1147 & 0.6061 & 0.3042 & -0.0173 & 0.0293 & 0.4040 \\
Factor 16 & 0.3362 & 1.8184 & 0.1525 & 0.1781 & 1.0645 & 0.1746 \\
Factor 17 & 0.9507 & 5.3376 & 0.0800 & 0.7202 & 4.3443 & 0.0942 \\
Factor 18 & 5.7249 & 11.4682 & 0.0639 & 4.9342 & 10.7183 & 0.0653 \\
Factor 19 & 0.1225 & 0.6743 & 0.2843 & -0.0103 & 0.0493 & 0.3732 \\
Factor 20 & -0.1281 & -0.9851 & 0.5210 & -0.2314 & -1.9506 & 0.7447 \\
Factor 21 & 0.0731 & 0.5597 & 0.2599 & -0.0539 & -0.3101 & 0.3486 \\
Factor 22 & 3.7633 & 5.9544 & 0.2400 & 3.2024 & 5.4859 & 0.2496 \\
Factor 23 & 0.0925 & 0.8389 & 0.2159 & -0.0368 & -0.2742 & 0.3663 \\
Factor 24 & 0.1004 & 0.6618 & 0.2489 & -0.0299 & -0.1014 & 0.3188 \\
Factor 25 & 0.1196 & 0.8420 & 0.2177 & -0.0129 & -0.0151 & 0.3896 \\
Factor 26 & 0.0068 & 0.1291 & 0.3761 & -0.1124 & -0.5531 & 0.4720 \\
Factor 27 & -0.0257 & -0.0189 & 0.4769 & -0.1411 & -0.6175 & 0.6157 \\
Factor 28 & 1.0541 & 3.0632 & 0.2612 & 0.8115 & 2.5491 & 0.2910 \\
Factor 29 & -0.0929 & -0.4082 & 0.4854 & -0.2004 & -1.0603 & 0.7017 \\
Factor 30 & 0.0756 & 0.4986 & 0.2876 & -0.0517 & -0.2093 & 0.4325 \\
Factor 31 & 1.1063 & 3.7826 & 0.2093 & 0.8576 & 3.1608 & 0.2277 \\
Factor 32 & -0.0405 & -0.1420 & 0.3815 & -0.1541 & -0.8470 & 0.5971 \\
Factor 33 & 0.0965 & 0.5860 & 0.3367 & -0.0333 & -0.0879 & 0.4423 \\
Factor 34 & -0.8139 & -11.0186 & 1.0000 & -0.8361 & -11.8528 & 0.9999 \\
Factor 35 & 0.2667 & 2.0504 & 0.3014 & 0.1169 & 0.9899 & 0.3517 \\
Factor 36 & 0.0136 & 0.1653 & 0.3583 & -0.1064 & -0.5321 & 0.4834 \\
Factor 37 & 0.1896 & 1.1592 & 0.2196 & 0.0489 & 0.3768 & 0.3050 \\
Factor 38 & 0.4755 & 2.5532 & 0.1403 & 0.3010 & 1.7520 & 0.1658 \\
Factor 39 & 0.0805 & 0.5837 & 0.2006 & -0.0474 & -0.2426 & 0.2925 \\
Factor 40 & 0.0683 & 0.4279 & 0.3222 & -0.0582 & -0.1955 & 0.3959 \\
\hline
\end{tabularx}
\begin{tablenotes}[flushleft]\footnotesize
\item Notes: This table shows In-Sample performance of factors in long-only portfolios, calculated by GPT. In-Sample means the backtesting period occurs before the cutoff date of GPT’s training data. It involves long positions in the top 10\% of futures by factor values and short positions in the bottom 10\%. Data covers the period from 2018/01/02 to 2023/04/07, with historical data from 2010 to 2021 used for reference. Symbol \* indicates results adjusted for transaction fees.
\end{tablenotes}
\end{threeparttable}
\end{table}

\begin{table}[H]
\centering
\begin{threeparttable}
\caption{In-Sample Alphas of Single Factor Portfolio}
\label{ISA}
\vspace{1em} 
\begin{tabular}{l@{\hskip 40pt}l@{\hskip 40pt}l@{\hskip 40pt}l@{\hskip 40pt}l@{\hskip 40pt}l}
\hline
\textbf{Factor} & \textbf{Long-Short} & \textbf{Long-Only} & \textbf{Factor} & \textbf{Long-Short} & \textbf{Long-Only} \\
\hline
factor1 & 0.2530*** & 0.5126*** & factor21 & 0.0260* & 0.0672 \\
factor2 & 0.0326** & 0.1884*** & factor22 & 0.3812*** & 1.1835*** \\
factor3 & 0.0412** & 0.1378 & factor23 & 0.0094 & 0.0751* \\
factor4 & 0.0670*** & 0.2171*** & factor24 & 0.0117 & 0.0837 \\
factor5 & 0.0042 & 0.0211 & factor25 & 0.0246 & 0.0717 \\
factor6 & 0.0075 & 0.0432 & factor26 & -0.0012 & 0.0049 \\
factor7 & -1.1335*** & -2.2236*** & factor27 & -0.0379* & -0.0528 \\
factor8 & -0.0040 & 0.0845 & factor28 & 0.2281*** & 0.5608*** \\
factor9 & 0.2551*** & 0.6112*** & factor29 & -0.0169 & -0.0700 \\
factor10 & -0.0309* & 0.0020 & factor30 & 0.0016 & 0.0500 \\
factor11 & -0.0466** & 0.0493 & factor31 & 0.2316*** & 0.5545*** \\
factor12 & 0.0081 & 0.0280 & factor32 & -0.0136 & -0.0252 \\
factor13 & 0.0911*** & 0.3804*** & factor33 & 0.0495*** & 0.0845 \\
factor14 & 0.0707*** & 0.2683*** & factor34 & -0.5960*** & -1.1976*** \\
factor15 & 0.0118 & 0.0726 & factor35 & 0.0164 & 0.1957*** \\
factor16 & 0.0468*** & 0.2307*** & factor36 & -0.0206 & 0.0349 \\
factor17 & 0.1939*** & 0.5257*** & factor37 & -0.0436*** & 0.1532*** \\
factor18 & 0.1450*** & 1.3927*** & factor38 & 0.1548*** & 0.2954*** \\
factor19 & 0.0050 & 0.0863 & factor39 & 0.0306** & 0.0602 \\
factor20 & -0.0565*** & -0.1032** & factor40 & -0.0015 & 0.0555 \\
\hline
\end{tabular}
\begin{tablenotes}[flushleft]\footnotesize
\item Notes: This table depicts the In-Sample alpha of each GPT-generated factor within long-short and long-only portfolios. Statistical significance is denoted as: *** for 1\% significance, ** for 5\% significance, and * for 10\% significance.
\end{tablenotes}
\end{threeparttable}
\end{table}

\begin{table}[H]
\centering
\begin{threeparttable}
\caption{In-Sample Performance of Factors: Multi-factor Portfolio}
\label{ISM}
\vspace{1em} 
\begin{tabularx}{\textwidth}{l *{6}{>{\centering\arraybackslash}X}}
\toprule
\multicolumn{7}{c}{\textbf{Static Multi-Factor Portfolio Returns}} \\
\midrule
\textbf{Strategy} & \textbf{Annualized return} & \textbf{Sharpe ratio} & \textbf{Max drawdown} & \textbf{Annualized return*} & \textbf{Sharpe ratio*} & \textbf{Max drawdown*} \\
\midrule
long-only & 0.6528 & 2.2600 & 0.2704 & 0.4574 & 1.7232 & 0.3144 \\
long-short & 0.8111 & 3.0218 & 0.1623 & 0.5971 & 2.4031 & 0.1786 \\
\midrule
\multicolumn{7}{>{\hsize=\textwidth\centering}X}{\textbf{Dynamic Multi-Factor Portfolio Returns - Referenced by Long-Short Returns}}\\
\midrule
\textbf{Strategy} & \textbf{Annualized return} & \textbf{Sharpe ratio} & \textbf{Max drawdown} & \textbf{Annualized return*} & \textbf{Sharpe ratio*} & \textbf{Max drawdown*} \\
\midrule
long-only & 0.6533 & 2.7674 & 0.1596 & 0.4578 & 2.0979 & 0.1849 \\
long-short & 1.3082 & 4.1289 & 0.0844 & 1.0357 & 3.5234 & 0.0945 \\
\midrule
\multicolumn{7}{>{\hsize=\textwidth\centering}X}{\textbf{Dynamic Multi-Factor Portfolio Returns - Referenced by Long-Only Returns}} \\
\midrule
\textbf{Strategy} & \textbf{Annualized return} & \textbf{Sharpe ratio} & \textbf{Max drawdown} & \textbf{Annualized return*} & \textbf{Sharpe ratio*} & \textbf{Max drawdown*} \\
\midrule
long-only & 0.5553 & 2.4939 & 0.2452 & 0.3599 & 1.7814 & 0.2914 \\
long-short & 0.7121 & 2.6781 & 0.2189 & 0.5304 & 2.1798 & 0.2471 \\
\bottomrule
\end{tabularx}
\begin{tablenotes}[flushleft]\footnotesize
\item Notes: This table presents In-Sample performance of multi-factor portfolios, which include both static and dynamic approaches as well as long-only and long-short strategies, using factors created by GPT. In the static approach, the sign (positive or negative) of each factor is determined based on their historical returns from 2010/01/04 to 2017/12/29. Conversely, the dynamic approach adjusts the sign of each factor daily, reflecting the cumulative returns of associated strategies, whether long-only or long-short. Symbol \* indicates results adjusted for transaction fees.
\end{tablenotes}
\end{threeparttable}
\end{table}

\begin{table}[H]
\centering
\begin{threeparttable}
\caption{In-Sample Alphas of Multi-factor Portfolio}
\label{ISAM}
\begin{tablenotes}[flushleft]\footnotesize
\begin{tabular}{l@{\hskip 80pt}c@{\hskip 80pt}c}
\toprule
\textbf{Factor} & \textbf{Long-Only Alpha} & \textbf{Long-Short Alpha}\\
\midrule
Static & 0.1073*** & 0.1330*** \\
Dynamic$_{\text{long-only base}}$ & 0.1001*** & 0.1468*** \\
Dynamic$_{\text{long-short base}}$ & 0.0907*** & 0.1248*** \\
\bottomrule
\end{tabular}
\item Notes: The table presents the alpha of static and dynamic multi-factor portfolios using factors generated by GPT in a long-short combination. ‘long-only base’ refers to using the factor’s long-only returns as a positive or negative reference, and ‘long-short base’ means referencing the factor’s long-short returns. *** denotes 1\% statistical significance, ** denotes 5\% statistical significance, and * indicates 10\% statistical significance.
\end{tablenotes}
\end{threeparttable}
\end{table}

For the dynamic portfolio referenced by long-short returns, the long-only strategy achieved an annual return of 0.6533, with a Sharpe ratio of 2.7674 and a maximum drawdown of 0.1596. This performance, while lower in absolute returns compared to the static model, demonstrates a reasonable balance between return and risk. The dynamic adjustment helps in responding to market conditions, thus optimizing the return profile while managing drawdown more effectively. The dynamic long-short portfolio, on the other hand, achieved an annual return of 1.3082, accompanied by an impressive Sharpe ratio of 4.1289 and a maximum drawdown of 0.0844. Compared to the long-only strategy, it mirrors the static portfolio model's trend of higher returns and lower risk. This demonstrates that the dynamic long-short strategy not only maintains its superior risk-adjusted returns but also significantly reduces drawdown.

When the dynamic multi-factor portfolio returns are benchmarked by long-only returns, the long-only strategy posts an annual return of 0.5553, with a Sharpe ratio of 2.4939 and a maximum drawdown of 0.2452. Meanwhile, the dynamic long-short strategy shows a return of 0.7121, a Sharpe ratio of 2.6781, and a maximum drawdown of 0.2189. This comparison reveals that when using long-only as the benchmark, both the static portfolio and the other dynamic portfolio exhibit lower returns and higher risk. This further underscores the robustness of the long-short investment strategy, as it consistently outperforms the long-only approach in terms of return and risk management. The dynamic long-short strategy demonstrates its superiority by achieving higher returns and better risk-adjusted performance. What's more, in the case of the multi-factor portfolio models, the impact of transaction fees becomes even less significant. 

Next, we examine the performance of multi-factor portfolio models relative to the IPCA benchmark, focusing exclusively on the long-short strategy to construct the portfolio models. Based on Table \ref{ISAM}, all multi-factor portfolio models achieve statistically significant in-sample alphas under the long-short strategy. The static model delivers an alpha of 0.1330***, highlighting its strong performance over the IPCA benchmark. The dynamic model based on long-only benchmarks achieves the highest alpha at 0.1468***, reflecting its superior adaptability to market changes. The dynamic model using long-short benchmarks also performs well, with an alpha of 0.1248***, effectively balancing return and risk through dynamic allocation.

\subsection{Out-of-Sample Performance}
In the preceding sections, we emphasized the use of a zero-shot approach in factor generation, which substantially mitigates the issue of forward-looking bias. To further examine the potential influence of GPT's training knowledge on the validity of the generated factors, we define the period following the cutoff date of GPT’s training data (April 2023) as the “Out-of-Sample” period, specifically spanning from 2023/05/04 to 2024/10/31. During this timeframe, we construct investment portfolios and conduct backtesting to evaluate whether the factors remain effective. This analysis serves to demonstrate that the performance of GPT-generated factors is not contingent upon the model’s pre-trained knowledge, thereby providing additional evidence of their robustness and reliability.

Apart from differences in the backtesting period, we employ the same methodology as earlier. Specifically, we construct both single-factor and multi-factor portfolios under various strategies to evaluate their annualized returns, Sharpe ratios, and maximum drawdowns. Table \ref{OOSLS} and \ref{OOSLO} present the performance of different single-factor portfolios under long-short and long-only strategies, respectively, while Table \ref{OOSA} reports the alpha results for single factor portfolios. 

From the results, we observe that the overall out-of-sample performance of the factors remains strong. Most factors continue to exhibit high annualized returns, elevated Sharpe ratios, and acceptable maximum drawdowns. Additionally, long-short strategies generally outperform long-only strategies, consistent with our previous findings, and the impact of transaction costs remains negligible. Furthermore, regarding in-sample and out-of-sample performance, we note a clear pattern: factors that performed well (poorly) in-sample tend to maintain their relative superiority (inferiority) out-of-sample compared to other factors. 

\begin{table}[htbp]
\centering
\begin{threeparttable}
\caption{Out-of-Sample Performance of Factors: Long-Short Portfolio}
\label{OOSLS}
\vspace{1em} 
\centering
\begin{tabularx}{\textwidth}{>{\raggedright\arraybackslash}p{2cm} >{\raggedright\arraybackslash}p{2cm} >{\raggedright\arraybackslash}p{2cm} >{\raggedright\arraybackslash}p{2cm} >{\raggedright\arraybackslash}p{2cm}>{\raggedright\arraybackslash}p{2cm}>{\raggedright\arraybackslash}p{2.5cm}}
\hline
\textbf{Factor} & \textbf{Annualized Return} & \textbf{Sharpe Ratio} & \textbf{Max Drawdown} & \textbf{Annualized Return*} & \textbf{Sharpe Ratio*} & \textbf{Max Drawdown*} \\
\hline
Factor 1  & 2.7739  & 7.0879  & 0.0740  & 2.3292  & 6.4261  & 0.0773  \\
Factor 2  & 0.2828  & 0.9743  & 0.2385  & 0.1310  & 0.5590  & 0.2515  \\
Factor 3  & -0.1433  & -0.7122  & 0.3600  & -0.2448  & -1.3724  & 0.4178  \\
Factor 4  & 0.7500  & 3.3514  & 0.0781  & 0.5432  & 2.6168  & 0.0814  \\
Factor 5  & -0.1258  & -0.6152  & 0.3970  & -0.2294  & -1.2833  & 0.4620  \\
Factor 6  & 0.4817  & 1.6384  & 0.2119  & 0.3065  & 1.1543  & 0.2335  \\
Factor 7  & -0.9947  & -28.8507  & 0.9993  & -0.9954  & -29.5534  & 0.9994  \\
Factor 8  & 0.5758  & 1.9943  & 0.1679  & 0.3895  & 1.4757  & 0.1922  \\
Factor 9  & 4.5350  & 8.0580  & 0.0393  & 3.8837  & 7.4743  & 0.0417  \\
Factor 10  & -0.2553  & -1.4740  & 0.4100  & -0.3436  & -2.1440  & 0.4979  \\
Factor 11  & 0.0878  & 0.4628  & 0.2638  & -0.0410  & -0.0476  & 0.2942  \\
Factor 12  & -0.0034  & 0.1056  & 0.3140  & -0.1214  & -0.4100  & 0.3738  \\
Factor 13  & 0.5272  & 1.7413  & 0.1592  & 0.3466  & 1.2620  & 0.2031  \\
Factor 14  & 1.0169  & 3.7248  & 0.0859  & 0.7787  & 3.0736  & 0.1090  \\
Factor 15  & 0.2061  & 1.1366  & 0.1465  & 0.0634  & 0.4322  & 0.1626  \\
Factor 16  & 0.8096  & 2.4976  & 0.0779  & 0.5958  & 1.9928  & 0.1121  \\
Factor 17  & 1.5724  & 8.3689  & 0.0273  & 1.2689  & 7.2624  & 0.0327  \\
Factor 18  & 0.8360  & 4.8589  & 0.0904  & 0.6191  & 3.8656  & 0.1049  \\
Factor 19  & 1.0093  & 2.6054  & 0.1118  & 0.7719  & 2.1599  & 0.1381  \\
Factor 20  & 1.0409  & 4.7372  & 0.0477  & 0.7999  & 3.9150  & 0.0652  \\
Factor 21  & 0.3861  & 1.7406  & 0.2035  & 0.2221  & 1.1075  & 0.2462  \\
Factor 22  & 15.1197  & 8.8804  & 0.0743  & 13.2302  & 8.4870  & 0.0767  \\
Factor 23  & -0.3224  & -1.2361  & 0.5109  & -0.4028  & -1.6836  & 0.5558  \\
Factor 24  & 0.0144  & 0.1825  & 0.1020  & -0.1057  & -0.9257  & 0.1860  \\
Factor 25  & 0.3278  & 1.2468  & 0.2078  & 0.1707  & 0.7482  & 0.2122  \\
Factor 26  & -0.4994  & -2.6483  & 0.6222  & -0.5588  & -3.1552  & 0.6825  \\
Factor 27  & 2.5225  & 4.3508  & 0.1460  & 2.1073  & 3.9308  & 0.1632  \\
Factor 28  & 2.9793  & 5.7651  & 0.1318  & 2.5105  & 5.2515  & 0.1401  \\
Factor 29  & -0.1405  & -1.3718  & 0.2581  & -0.2424  & -2.5576  & 0.3536  \\
Factor 30  & -0.2620  & -1.3248  & 0.4215  & -0.3495  & -1.9194  & 0.5108  \\
Factor 31  & 1.3855  & 3.2508  & 0.2576  & 1.1039  & 2.8009  & 0.2618  \\
Factor 32  & -0.4103  & -2.1909  & 0.5322  & -0.4803  & -2.7423  & 0.6013  \\
Factor 33  & 0.9278  & 2.8737  & 0.0572  & 0.7000  & 2.3445  & 0.0825  \\
Factor 34  & 22.1293  & 13.5809  & 0.0266  & 19.4221  & 13.0439  & 0.0276  \\
Factor 35  & 0.2372  & 1.1044  & 0.3035  & 0.0908  & 0.5119  & 0.3520  \\
Factor 36  & 0.0444  & 0.3214  & 0.1688  & -0.0792  & -0.3318  & 0.1902  \\
Factor 37  & -0.2681  & -2.7105  & 0.3538  & -0.3548  & -3.8284  & 0.4578  \\
Factor 38  & 3.4112  & 5.4908  & 0.0711  & 2.8917  & 5.0371  & 0.0826  \\
Factor 39  & 1.4073  & 3.4661  & 0.0574  & 1.1231  & 2.9876  & 0.0626  \\
Factor 40  & -0.1702  & -1.1562  & 0.2310  & -0.2685  & -1.9883  & 0.3549  \\
\hline
\end{tabularx}
\begin{tablenotes}[flushleft]\footnotesize
\item Notes: This table shows Out-of-Sample performance of factors in long-short portfolios, calculated by GPT. Out-of-Sample means the backtesting period occurs after the cutoff date of GPT’s training data. It involves long positions in the top 10\% of futures by factor values and short positions in the bottom 10\%. Data covers the period from 2023/05/04 to 2024/10/31. Symbol * indicates results adjusted for transaction fees.
\end{tablenotes}
\end{threeparttable}
\end{table}

\begin{table}[htbp]
\centering
\begin{threeparttable}
\caption{Out-of-Sample Performance of Factors: Long-Only Portfolio}
\label{OOSLO}
\vspace{1em} 
\centering
\begin{tabularx}{\textwidth}{>{\raggedright\arraybackslash}p{2cm} >{\raggedright\arraybackslash}p{2cm} >{\raggedright\arraybackslash}p{2cm} >{\raggedright\arraybackslash}p{2cm} >{\raggedright\arraybackslash}p{2cm}>{\raggedright\arraybackslash}p{2cm}>{\raggedright\arraybackslash}p{2.5cm}}
\hline
\textbf{Factor} & \textbf{Annualized Return} & \textbf{Sharpe Ratio} & \textbf{Max Drawdown} & \textbf{Annualized Return*} & \textbf{Sharpe Ratio*} & \textbf{Max Drawdown*} \\
\hline
Factor 1  & 1.1613  & 3.8061  & 0.1023  & 0.9061  & 3.2016  & 0.1055  \\
Factor 2  & -0.0006  & 0.1408  & 0.2996  & -0.1190  & -0.3120  & 0.3384  \\
Factor 3  & -0.0966  & -0.3657  & 0.3626  & -0.2036  & -0.9542  & 0.4182  \\
Factor 4  & 0.6109  & 2.4274  & 0.1142  & 0.4205  & 1.8133  & 0.1398  \\
Factor 5  & -0.1167  & -0.5245  & 0.3410  & -0.2213  & -1.1604  & 0.4012  \\
Factor 6  & 0.0430  & 0.3148  & 0.1900  & -0.0805  & -0.3384  & 0.2622  \\
Factor 7  & -0.9190  & -14.5347  & 0.9702  & -0.9286  & -15.2715  & 0.9750  \\
Factor 8  & 0.5245  & 1.7158  & 0.1971  & 0.3443  & 1.2431  & 0.2166  \\
Factor 9  & 1.5701  & 4.6859  & 0.0757  & 1.2668  & 4.0751  & 0.0880  \\
Factor 10  & -0.1453  & -1.0131  & 0.2559  & -0.2465  & -1.8844  & 0.3572  \\
Factor 11  & 0.1772  & 0.7608  & 0.2297  & 0.0379  & 0.2725  & 0.2593  \\
Factor 12  & -0.0291  & -0.0813  & 0.2075  & -0.1441  & -0.8020  & 0.3251  \\
Factor 13  & 0.3253  & 1.0532  & 0.2814  & 0.1685  & 0.6521  & 0.3073  \\
Factor 14  & 0.6367  & 2.5139  & 0.1539  & 0.4432  & 1.8973  & 0.1761  \\
Factor 15  & 0.1910  & 0.9219  & 0.1788  & 0.0501  & 0.3342  & 0.2084  \\
Factor 16  & 0.1042  & 0.7462  & 0.1468  & -0.0265  & -0.1089  & 0.1795  \\
Factor 17  & 0.7932  & 4.4036  & 0.0529  & 0.5813  & 3.4689  & 0.0684  \\
Factor 18  & 3.6893  & 10.7747  & 0.0348  & 3.1372  & 9.9046  & 0.0381  \\
Factor 19  & 0.2817  & 1.2840  & 0.2117  & 0.1300  & 0.6851  & 0.2442  \\
Factor 20  & 0.5953  & 2.9118  & 0.1096  & 0.4067  & 2.1489  & 0.1446  \\
Factor 21  & 0.1982  & 1.0077  & 0.1696  & 0.0564  & 0.3749  & 0.2562  \\
Factor 22  & 3.8655  & 6.4786  & 0.1155  & 3.2926  & 5.9741  & 0.1292  \\
Factor 23  & -0.3717  & -1.9662  & 0.5090  & -0.4462  & -2.5318  & 0.5792  \\
Factor 24  & 0.1124  & 0.8583  & 0.1886  & -0.0193  & -0.0778  & 0.2198  \\
Factor 25  & 0.0111  & 0.1522  & 0.2506  & -0.1086  & -0.5942  & 0.2882  \\
Factor 26  & -0.4570  & -2.2390  & 0.5808  & -0.5215  & -2.7295  & 0.6479  \\
Factor 27  & 0.7771  & 2.7226  & 0.2310  & 0.5671  & 2.1504  & 0.2570  \\
Factor 28  & 0.9957  & 3.0914  & 0.1445  & 0.7599  & 2.5493  & 0.1602  \\
Factor 29  & -0.1200  & -0.7527  & 0.2769  & -0.2243  & -1.5704  & 0.3551  \\
Factor 30  & -0.2153  & -1.0117  & 0.3258  & -0.3084  & -1.5958  & 0.4259  \\
Factor 31  & 0.3334  & 1.2720  & 0.2420  & 0.1757  & 0.7717  & 0.2466  \\
Factor 32  & -0.3442  & -1.7217  & 0.4589  & -0.4220  & -2.2717  & 0.5391  \\
Factor 33  & 0.1710  & 1.0212  & 0.1486  & 0.0324  & 0.2725  & 0.2065  \\
Factor 34  & 3.4093  & 8.9353  & 0.0481  & 2.8900  & 8.1857  & 0.0505  \\
Factor 35  & 0.0558  & 0.5868  & 0.2031  & -0.0692  & -0.6578  & 0.2818  \\
Factor 36  & -0.0880  & -0.4001  & 0.2317  & -0.1961  & -1.0751  & 0.3403  \\
Factor 37  & 0.0184  & 0.1983  & 0.1489  & -0.1022  & -0.6741  & 0.2146  \\
Factor 38  & 0.7262  & 3.7776  & 0.0752  & 0.5222  & 2.9235  & 0.0991  \\
Factor 39  & 0.2300  & 1.6244  & 0.1040  & 0.0845  & 0.6764  & 0.1392  \\
Factor 40  & 0.0254  & 0.2365  & 0.1532  & -0.0960  & -0.5533  & 0.2025  \\
\hline
\end{tabularx}
\begin{tablenotes}[flushleft]\footnotesize
\item Notes: This table shows Out-of-Sample performance of factors in long-only portfolios, calculated by GPT. Out-of-Sample means the backtesting period occurs after the cutoff date of GPT’s training data. It involves long positions in the top 10\% of futures by factor values and short positions in the bottom 10\%. Data covers the period from 2023/05/04 to 2024/10/31. Symbol * indicates results adjusted for transaction fees.
\end{tablenotes}
\end{threeparttable}
\end{table}

\begin{table}[htbp]
\centering
\begin{threeparttable}
\caption{Out-of-Sample Alphas of Single Factor Portfolio}
\label{OOSA}
\vspace{1em} 
\begin{tabular}{l@{\hskip 40pt}l@{\hskip 40pt}l@{\hskip 40pt}l@{\hskip 40pt}l@{\hskip 40pt}l}
\hline
\textbf{Factor} & \textbf{Long-Short} & \textbf{Long-Only} & \textbf{Factor} & \textbf{Long-Short} & \textbf{Long-Only} \\
\hline
Factor1 & 0.3200*** & 0.5953*** & Factor21 & 0.0969** & 0.2108** \\
Factor2 & 0.0598 & 0.2185 & Factor22 & 0.7067*** & 1.1717*** \\
Factor3 & -0.0347 & 0.0433 & Factor23 & -0.0749 & -0.2268* \\
Factor4 & 0.1360*** & 0.4508*** & Factor24 & 0.0097 & 0.1674** \\
Factor5 & -0.0207 & 0.0630 & Factor25 & 0.0652 & 0.0658 \\
Factor6 & 0.1008* & 0.1242 & Factor26 & -0.1723*** & -0.2779** \\
Factor7 & -1.2795*** & -1.4237*** & Factor27 & 0.3286*** & 0.5444*** \\
Factor8 & 0.1037** & 0.4701*** & Factor28 & 0.3453*** & 0.5814*** \\
Factor9 & 0.4152*** & 0.7344*** & Factor29 & -0.0376* & 0.0223 \\
Factor10 & -0.0591 & -0.0222 & Factor30 & -0.0809* & -0.0152 \\
Factor11 & 0.0067 & 0.2841** & Factor31 & 0.2100*** & 0.3361** \\
Factor12 & 0.0270 & 0.0993 & Factor32 & -0.1260** & -0.1704 \\
Factor13 & 0.1012* & 0.4356*** & Factor33 & 0.1644*** & 0.2367*** \\
Factor14 & 0.1663*** & 0.4612*** & Factor34 & 0.7636*** & 1.0408*** \\
Factor15 & 0.0385 & 0.2688** & Factor35 & 0.0586 & 0.1019* \\
Factor16 & 0.1566*** & 0.1533* & Factor36 & 0.0178 & 0.0335 \\
Factor17 & 0.2390*** & 0.4603*** & Factor37 & -0.0785*** & 0.1263 \\
Factor18 & 0.1449*** & 1.0473*** & Factor38 & 0.3793*** & 0.4240*** \\
Factor19 & 0.1612*** & 0.3126*** & Factor39 & 0.2290*** & 0.1998*** \\
Factor20 & 0.1900*** & 0.3704*** & Factor40 & -0.0458 & 0.1191 \\
\hline
\end{tabular}
\begin{tablenotes}[flushleft]\footnotesize
\item Notes: This table depicts the Out-of-Sample alpha of each GPT-generated factor within long-short and long-only portfolios. Statistical significance is denoted as: *** for 1\% significance, ** for 5\% significance, and * for 10\% significance.
\end{tablenotes}
\end{threeparttable}
\end{table}

In the analysis of multi-factor portfolio performance from table \ref{OOSM} and \ref{OOSAM}, we find that multi-factor portfolios continue to demonstrate robust results across all strategies. Their performance shows no significant differences compared to the in-sample period and, in some cases, even surpasses it. Furthermore, dynamic multi-factor portfolios consistently outperform their static counterparts, with notable improvements in Sharpe ratios and reductions in maximum drawdowns. These findings align closely with the in-sample results, reinforcing the advantages of dynamic factor weighting in capturing market conditions and enhancing portfolio stability.

The analysis presented in this section underscores the necessity of evaluating the out-of-sample performance of GPT-generated factors. By rigorously testing these factors in a period beyond GPT's training data cutoff, we address potential concerns regarding their dependence on pre-trained knowledge and validate their adaptability to new market conditions. The consistent results across various strategies, coupled with the robust performance of both single-factor and multi-factor portfolios, highlight the resilience and effectiveness of GPT-generated factors. These findings not only reinforce the credibility of GPT as a tool for financial factor generation but also emphasize its potential to create stable and reliable investment strategies, even in dynamic and evolving market environments.

\begin{table}[H]
\centering
\begin{threeparttable}
\caption{Out-of-Sample Performance of Factors: Multi-factor Portfolio}
\label{OOSM}
\vspace{1em} 
\begin{tabularx}{\textwidth}{l *{6}{>{\centering\arraybackslash}X}}
\toprule
\multicolumn{7}{c}{\textbf{Static Multi-Factor Portfolio Returns}} \\
\midrule
\textbf{Strategy} & \textbf{Annualized return} & \textbf{Sharpe ratio} & \textbf{Max drawdown} & \textbf{Annualized return*} & \textbf{Sharpe ratio*} & \textbf{Max drawdown*} \\
\midrule
long-only & 0.8473 & 2.4147 & 0.1836 & 0.6290 & 1.9468 & 0.2059 \\
long-short & 1.5955 & 3.4423 & 0.1489 & 1.2892 & 3.0069 & 0.1722 \\
\midrule
\multicolumn{7}{>{\hsize=\textwidth\centering}X}{\textbf{Dynamic Multi-Factor Portfolio Returns - Referenced by Long-Short Returns}}\\
\midrule
\textbf{Strategy} & \textbf{Annualized return} & \textbf{Sharpe ratio} & \textbf{Max drawdown} & \textbf{Annualized return*} & \textbf{Sharpe ratio*} & \textbf{Max drawdown*} \\
\midrule
long-only & 0.7690 & 2.7765 & 0.0972 & 0.3629 & 1.4100 & 0.1108 \\
long-short & 1.2948 & 3.6335 & 0.1072 & 0.8075 & 2.3292 & 0.1802 \\
\midrule
\multicolumn{7}{>{\hsize=\textwidth\centering}X}{\textbf{Dynamic Multi-Factor Portfolio Returns - Referenced by Long-Only Returns}} \\
\midrule
\textbf{Strategy} & \textbf{Annualized return} & \textbf{Sharpe ratio} & \textbf{Max drawdown} & \textbf{Annualized return*} & \textbf{Sharpe ratio*} & \textbf{Max drawdown*} \\
\midrule
long-only & 1.0309 & 3.5529 & 0.0902 & 0.7900 & 2.9348 & 0.0958 \\
long-short & 2.1175 & 4.0800 & 0.0981 & 1.7070 & 3.5962 & 0.1056 \\
\bottomrule
\end{tabularx}
\begin{tablenotes}[flushleft]\footnotesize
\item Notes: This table presents Out-of-Sample performance of multi-factor portfolios, which include both static and dynamic approaches as well as long-only and long-short strategies, using factors created by GPT. In the static approach, the sign (positive or negative) of each factor is determined based on their historical returns from 2023/05/04 to 2023/12/31. Conversely, the dynamic approach adjusts the sign of each factor daily, reflecting the cumulative returns of associated strategies, whether long-only or long-short. Symbol * indicates results adjusted for transaction fees.
\end{tablenotes}
\end{threeparttable}
\end{table}

\begin{table}[htbp]
\centering
\begin{threeparttable}
\caption{Out-of-Sample Alphas of Multi-factor Portfolio}
\label{OOSAM}
\begin{tabular}{l@{\hskip 80pt}c@{\hskip 80pt}c}
\toprule
\textbf{Factor} & \textbf{Long-Only Alpha} & \textbf{Long-Short Alpha}\\
\midrule
Static & 0.2911*** & 0.3022*** \\
Dynamic$_{\text{long-only base}}$ & 0.1578*** & 0.1754*** \\
Dynamic$_{\text{long-short base}}$ & 0.3244*** & 0.3418*** \\
\bottomrule
\end{tabular}
\begin{tablenotes}[flushleft]\footnotesize
\item Notes: The table presents the Out-of-Sample alpha of static and dynamic multi-factor portfolios using factors generated by GPT in a long-short combination. ‘long-only base’ refers to using the factor’s long-only returns as a positive or negative reference, and ‘long-short base’ means referencing the factor’s long-short returns. *** denotes 1\% statistical significance, ** denotes 5\% statistical significance, and * indicates 10\% statistical significance.
\end{tablenotes}
\end{threeparttable}
\end{table}

\section{Anatomy of GPT-generated Factors}
\subsection{Decoding GPT's Factor Generation Process}
Driven by a desire to understand how GPT crafts these elements, after we comprehensively anatomized all of the codes provided by GPT for each factor, we conjecture that the following procedures were executed by GPT to do the job of factor generation.   

(1) Selection of Relevant Market Features: The choice of features (like price, volume, volatility) is crucial and is based on financial theory and practical significance. Each feature captures a different aspect of market behavior like price-based features (open, high, low, close), volume indicators, volatility indicators and other metrics like basis, futures premium, and discount relate to market sentiment and expectations.

(2) Calculation of Derived Statistics and Normalization: To transform raw data into insightful indicators, derived statistics such as moving averages and differences between short-term and long-term averages are calculated to identify trends. Subsequently, normalization techniques, like converting data to Z-scores, standardize these indicators across varying scales and distributions. 

(3) Composite Indicator Construction: Combining individual indicators (like momentum, volatility, sentiment) into a composite indicator like equal weighting and custom Weighting.

(4) Adaptation and Contextualization: Factors are often adapted to the specific characteristics of the data set or the particular needs of the analysis like rolling Windows for calculations can vary in size to capture different market dynamics and adjustments for Market Specifics like different trading volumes or market cap sizes.

(5) Analytical Objectivity: While developing factors, maintaining analytical objectivity is crucial, which means testing for Statistical Significance of the indicators and validation against historical data to check the factor’s predictive power and robustness.

(6) Continuous Evaluation and Iteration: Market conditions evolve, and so should the factors. Continuous back-testing and adjustment of the factors ensure they remain relevant and effective in capturing market dynamics.


We also directly ask GPT to elucidate its factor generation process, and it returns: “When I create financial factors, I start by examining the market data available, such as price, volume, and derived metrics like volatility. I focus on identifying patterns or trends that can provide insights into market behavior. Using statistical methods, I calculate various indicators — for instance, momentum from moving averages or volatility from price fluctuations. I then normalize these metrics to ensure consistency across different datasets and scales, which helps in comparing and merging them into composite indicators. Finally, I combine these normalized indicators using methods like simple addition or weighted averages to form a single factor that reflects multiple aspects of market dynamics. This process allows me to distill complex market information into a manageable and actionable form, aiming to reveal deeper market insights and potential investment opportunities.” We believe this answer is essentially consistent with our conjecture above.

\subsection{Decomposing the Elements of Factors}
In the realm of financial analysis, the effectiveness of a factor largely hinges on its constituent components. Each component serves a specific function, working in concert to distill complex market behaviors into actionable insights. This section aims to unravel the intricacies of each factor, providing a granular view of how individual elements are selected, calculated, and integrated. By dissecting these components, we gain not only a deeper understanding of what drives each factor but also how they collectively influence investment decisions and risk assessments. As we proceed, we will explore the theoretical underpinnings and practical implementations that make up the backbone of our financial indicators, shedding light on their predictive power and relevance in today’s dynamic market environment.

We provide a detailed description of each factor in Table B1, Online Appendix B, categorizing them by their functional components—momentum, volatility, sentiment, liquidity, and others. For each factor listed, the appendix describes the computational methods and the data sources used, providing a thorough understanding of the mechanics underlying each indicator. By breaking down the calculation methodologies and the rationale behind each component, the factors reveal diverse aspects of market behavior. For instance, the combination of short-term and long-term moving averages in the Innovative Momentum Volatility Sentiment Indicator (IMVSI) captures both immediate and overarching trends, offering insights into potential momentum shifts before they become evident to the market at large. Further, the Adjusted Liquidity Oscillation Weighted Sentiment (ALOWS) leverages volume and price oscillations to assess market liquidity under varying conditions, making it particularly valuable in fast-moving or illiquid markets where traditional indicators might lag or fail. These detailed analyses underscore the importance of each factor's sensitivity to specific market inputs—be it price movements, volume changes, or broader economic indicators like futures premiums and discounts. 

In the following sections, we will explore the commonalities among high-performing (annualized return > 1, base on long-short returns), low-performing (annualized return < -0.1), and average-performing factors (-0.03 < annualized return < 0.03). We find that high-performing factors (e.g., 1, 9, 17, 18, 22, 28, 31) achieve strong results by integrating price, volume, and sentiment with momentum or volatility measures—often using multi-period moving averages or standard deviations—to capture nuanced market dynamics. In contrast, low-performing factors (e.g., 7, 10, 11, 20, 29, 32, 34, 37) rely on basic price-volume metrics and simple normalization, lacking sentiment and volatility integration, making them less adaptive. Average-performing factors (e.g., 5, 6, 8, 12, 23, 24, 26, 40) focus on short- to medium-term price-volume relationships but miss deeper trends due to simpler metrics and limited integration of momentum, volatility asymmetry, or multi-period dynamics—offering stability but inferior performance in complex conditions.

In addition, our observation highlights an important and intriguing point: high-performing factors typically avoid using basis as a core metric, while low- and average-performing ones frequently incorporate it. This suggests that GPT may not fully grasp futures market dynamics, where basis is key, and instead leans toward stock market mechanisms, relying on price and volume rather than futures-specific indicators. This highlights a gap in GPT's understanding of futures and underscores the need for more tailored approaches in futures factor modeling.

\subsection{Economic Insights}
In delving into the economic ramifications of certain financial indicators, this analysis accentuates four particularly enlightening factors: Factor 1 (IMVSI), Factor 6 (MRSI), Factor 7 (FSI), Factor 10 (MMLI) and Factor 34 (FMAT). Each factor is selected for its distinctive ability to shed light on specific market dynamics, facilitating a deeper understanding of momentum, sentiment, volatility, and overall market conditions.

Factor 1 plays a vital role due to its comprehensive integration of market momentum, volatility, and sentiment. By simultaneously assessing the speed and stability of price movements along with prevailing investor sentiment, this factor provides economically meaningful insights that help identify potential trend reversals or confirm ongoing market directions. This makes it particularly valuable for traders and portfolio managers seeking timely signals for market entry or exit. 

Factor 6 focuses on detecting extreme price movements and volume anomalies, enhanced by sentiment analysis derived from the correlation between price changes and trading volume. This combination enables the factor to effectively signal potential market peaks and troughs, making it especially useful for contrarian strategies or those aiming to capture trend reversals. 

Factor 7 offers a nuanced view of market expectations by analyzing the relationship between futures and spot prices—specifically through the basis. Its significance lies in revealing discrepancies that may reflect underlying market tensions or anticipation of major economic developments, offering traders and analysts a powerful tool for gauging sentiment and forecasting price movements. 

Factor 10 emphasizes market momentum and liquidity, computing a momentum score based on the 7-day rolling average of daily price changes and assessing liquidity via the volume-to-traded amount ratio. This dual focus provides a real-time assessment of both the strength and sustainability of price trends, helping market participants determine whether current movements are supported by sufficient trading activity—an essential consideration for evaluating trend durability and managing risk, particularly in volatile conditions. 

Finally, Factor 34 synthesizes daily price dynamics, basis, and liquidity indicators to deliver a holistic assessment of the futures market environment. Its economic value is rooted in its ability to uncover deeper market trends and liquidity conditions that may not be apparent through conventional analysis, thereby supporting more informed long-term investment decisions. Together, these factors offer distinct yet complementary insights into market behavior, spanning momentum, sentiment, valuation discrepancies, and liquidity.

Collectively, these factors not only enhance the understanding of specific market aspects but also contribute to a multi-faceted comprehension of market behavior. This is essential for effectively navigating the complexities of today’s financial environments. By integrating these indicators into cohesive trading or investment strategies, financial professionals can improve risk management and optimize returns, leveraging these insights to address modern market challenges.

\section{Robustness}
\subsection{Different LLMs Comparison}
It is natural to wonder whether other LLMs, rather than GPT, can also generate novel factors. Therefore, we will compare the performance of different LLMs in generating factors in this section. With the rapid advancement of natural language processing technologies, various LLMs, such as GPT-4o, Copilot, and Claude 3, have shown significant advantages in handling and generating complex data. However, different models may exhibit varying levels of effectiveness and performance when tasked with specific objectives. Therefore, we will evaluate several representative LLMs, assessing their performance in factor generation, accuracy, efficiency, and applicability. Table C1, Online Appendix C, summarizes factor performance comparison of different LLMs. 

GPT-4o, the latest iteration, shows slight improvements over its predecessor. These enhancements are evident in its performance metrics, although the differences are not substantial. GPT-4o maintains a consistent positive annual return and impressive Sharpe ratios, indicating a good balance between return and risk. The moderate maximum drawdown values highlight GPT-4o's ability to manage losses effectively.

Copilot, developed by Microsoft, represents another major player in the realm of LLMs. However, its performance in generating factors is less stellar compared to GPT-4o. The probability of encountering negative returns with Copilot is relatively high, reflecting its overall lower performance stability. While there are instances of high annual returns, these are overshadowed by more frequent negative outcomes. The Sharpe ratios for Copilot models vary widely, indicating inconsistency in balancing risk and return. Additionally, higher maximum drawdown values suggest greater vulnerability to significant losses.

Claude, developed by the American tech company Anthropic, presents a mixed bag in terms of performance. This model tends to exhibit less stability, with results often showing extreme variations. While Claude can achieve exceptionally high returns in some scenarios, it also runs the risk of severe losses. This volatility is reflected in the wide range of Sharpe ratios and significant maximum drawdown values, indicating that while the potential for high reward exists, so does the potential for high risk. This characteristic makes Claude less predictable and more suitable for high-risk, high-reward investment strategies.



Overall, the comparison highlights the strengths and weaknesses of each model. GPT-4o remains a reliable and consistent performer with moderate improvements. Copilot, despite its promising technology from Microsoft, tends to yield less consistent results with a higher likelihood of negative returns. Claude, while capable of extreme performance, often lacks the stability needed for more conservative investment strategies. This analysis provides valuable insights into the optimal use cases for each of these LLMs, guiding their application based on desired risk tolerance and performance expectations.

\subsection{Parameter Temperature Variations}
To evaluate the robustness and consistency of the financial factors generated by GPT—whose outputs can vary due to the "temperature" parameter controlling response randomness—we conducted an experiment across ten independent dialogue conversations, each generating five factors, allowing us to assess both average performance and variability following the methodology in Section 6. The results, summarized in table C2 and table C3, Online Appendix, provide a comprehensive view of the factors’ performance across multiple runs.

Table C2 presents the performance metrics for the long-short portfolios constructed from the GPT-generated factors. The results indicate significant variability in the performance of the GPT-generated factors. For instance, Factor1-2 (i.e., the second factor from the first conversation window. This notation applies in the same way hereafter.) stands out with an exceptionally high annualized return of 24.3286 and a Sharpe ratio of 16.4453, coupled with a minimal maximum drawdown of 0.0436. Such extraordinary performance suggests that this factor captured unique and highly profitable market conditions. In contrast, factors such as Factor2-2 and Factor3-3 demonstrate poor performance, with negative annualized returns of -0.5450 and -0.5642, respectively, and large maximum drawdowns exceeding 0.64. These results highlight the inherent risk and potential volatility associated with some GPT-generated factors.

Factors such as Factor5-3 and Factor7-3 also exhibit impressive performance, with annualized returns of -0.0449 and -0.1455, respectively. Their Sharpe ratios of 0.2626 and 0.3830 indicate moderate risk-adjusted returns, further validated by their relatively low maximum drawdowns of 0.2079 and 0.3830. Interestingly, several factors show moderate performance with both positive and negative returns. For example, Factor6-1 achieved an annualized return of 13.2776 and a Sharpe ratio of 13.8652, demonstrating a balanced performance with a modest maximum drawdown of 0.0330. Conversely, Factor6-4 shows a negative return of -0.4757 and a significant maximum drawdown of 0.5878, emphasizing the variability and unpredictability of some factors.

Table C3 further presents the performance metrics for the long-only portfolios. The results indicate a wide range of performance among the GPT-generated factors. Among the top-performing factors, Factor5-1 stands out due to its ability to deliver strong returns of 2.6920 while maintaining low risk, as evidenced by its Sharpe ratio of 7.3562 and maximum drawdown of 0.0753. This suggests that the factor captured particularly favorable market conditions. Conversely, factors like Factor2-2 and Factor3-3 demonstrate poor performance, with negative annualized returns of -0.3736 and -0.4775, respectively, and significant maximum drawdowns, indicating high risk and volatility. Mid-tier performers such as Factor8-3 and Factor10-3 demonstrate consistent value, with moderate annualized returns and Sharpe ratios indicating effective risk management. Meanwhile, Factor6-1 achieves notable success, balancing higher returns with manageable risks, whereas Factor4-1 adopts a more conservative profile, favoring lower volatility and steadier growth. Together, these factors provide a comprehensive perspective on the diverse applications and potential outcomes of GPT-generated portfolios under varying market scenarios.

In summary, this subsection evaluated the robustness and consistency of GPT-generated factors in both long-short and long-only portfolios across multiple dialogue sessions. The results indicate that different sessions, controlled by the temperature parameter, do not significantly affect the performance of GPT-generated factor models. Factors like Factor3-4 and Factor5-1 consistently showed exceptional returns in both strategies, underscoring the stability of GPT in generating reliable financial models. These findings demonstrate that the variability introduced by different dialogues does not substantially impact the robustness of GPT-generated factors, affirming their effectiveness across varying conditions.

\subsection{Prompt Engineering}
The effectiveness of GPT-generated financial factors heavily depends on prompt engineering, as the formulation of input prompts directly influences the quality and performance of the output. This analysis examines how varying prompt complexity—through simplified versus more intricate prompts—affects the robustness, stability, and overall performance of the generated factor models. By evaluating these variations, we aim to identify optimal prompt design strategies that enhance the reliability and practical utility of GPT in financial modeling.

\subsubsection{Simplified Prompts}
In this subsection, we investigate the effects of using simplified prompts on the performance of GPT-generated factor models. By reducing the complexity and specificity of the prompts provided to GPT, we aim to understand how minimal guidance impacts the quality and effectiveness of the generated financial factors. Specifically, the simplified prompts do not include instructions to establish a specific identity for GPT or to ensure the innovation and originality of the factors. Instead, the prompts focus solely on generating factors based on daily futures market data. By simplifying the prompt in this manner, we aim to assess whether the absence of explicit guidance affects the performance and robustness of the factors generated by GPT. The modified prompt used in this analysis is as follows:

“I need you to generate factors for me. You have panel data for many futures stored as a dataframe named ‘futures\_data’. I require you to provide me with a Python program that can generate a new column as a factor. The data columns include futuresname, date, basis, futures premium and discount, spot, open, high, low, close, volume, and amount.”

The performance of factors generated from simplified prompts in Table C4, Online Appendix C, shows considerable variability across both long-short and long-only portfolios. In the long-short setup, while factors like Simple7 (9.35\% return, Sharpe ratio 12.36) and Simple6 (9.42\% return, Sharpe ratio 9.04) achieved strong results with moderate drawdowns, many others—including Simple1, Simple2, Simple5 (–0.18\% return, Sharpe –0.81, 30.29\% drawdown), Simple8, and Simple10—produced negative returns and low or negative Sharpe ratios, indicating poor risk-adjusted performance. Similarly, in long-only portfolios, Simple15 delivered solid returns (1.47\%, Sharpe 5.87) with low drawdown (6.17\%), and Simple6 achieved 2.48\% return (Sharpe 5.47), but several factors such as Simple1 (–0.04\%, Sharpe –0.07) and Simple2 (–0.11\%, Sharpe 0.61) underperformed with higher drawdowns. Overall, these results suggest that simplified prompts tend to generate factors with inconsistent performance, higher risk, and lower average returns compared to the original models.

This analysis demonstrates that simplifying the prompts provided to GPT has a significant impact on the performance and robustness of the generated financial factors. Specifically, the simplified prompts led to factors that, on average, exhibited lower returns and higher risks. The maximum drawdowns in these factors were generally higher compared to those generated with more detailed and specific prompts. Moreover, due to the lack of constraints on the originality of the factors, GPT tended to output well-known technical indicators such as RSI and MACD. This overlap with existing research and practices underscores the importance of prompt engineering in guiding GPT to generate innovative and unique factors.

\subsubsection{Specific Prompts}
In this subsection, we investigate the impact of complex prompts on the performance of GPT-generated factor models. By providing GPT with more detailed and specific instructions, we aim to understand how increased complexity in prompt engineering affects the robustness and effectiveness of the generated factors.

Specifically, we initiated four distinct dialogue sessions, each with a unique focus to generate factors based on different analytical frameworks:

1. Trend-based Analysis: Prompts designed to generate factors that capture market trends.

2. Statistical Models: Prompts aimed at producing factors with intricate statistical formulations.

3. Stochastic Modeling: Prompts structured to create factors using stochastic modeling analysis techniques.

4. Behavioral Finance: Prompts crafted to develop factors related to behavioral finance principles.

For each of these analytical frameworks, we opened a new dialogue window with GPT and generated 5 new factors that satisfied the specific requirements of the prompts. Following the generation of these factors, we constructed single-factor models to evaluate their performance rigorously.

Table C5, Online Appendix C, summarizes the performance of factors generated using complex prompts across four categories: trend-based analysis, mathematically complex models, Fourier analysis, and behavioral finance, measured by annualized return, Sharpe ratio, and maximum drawdown. Trend-based factors show high variability—Trend2 achieves strong performance with a 0.723 return and 2.288 Sharpe ratio with low drawdown (0.1372), while Trend3 and Trend4 underperform. Mathematically complex factors also vary: Statistical1 and Statistical5 achieve high returns (13.6568 and 33.9494) and solid Sharpe ratios, whereas Statistical2 and Statistical3 show inconsistent results, highlighting the importance of careful model selection and validation.

One notable observation is the presence of outlier results, such as the extraordinarily high returns for Trend1 and Trend5 in long-short portfolio. These anomalies suggest that while some factors may perform exceptionally well under certain conditions, they may not be reliable or replicable in other scenarios. The extreme variability in performance highlights the instability of factors generated under highly specific and complex prompts. This variability and the occurrence of outliers indicate that overly specific prompts might constrain GPT’s ability to generate robust and consistently performing factors. By narrowing the focus too much, the model may produce factors that are highly optimized for certain data sets or market conditions but fail to generalize well across different scenarios.

Long-only portfolios show more stable performance than long-short ones, with consistent returns and lower drawdowns. Behavior4, Behavior2, and Trend2 deliver solid returns and Sharpe ratios with minimal drawdowns. Statistical1 and Statistical5 achieve strong returns (2.4709, 4.7083) and high Sharpe ratios, though Statistical3 and Statistical4 vary, emphasizing the need for validation. Stochastic factors are mixed: Stochastic2 performs modestly, while Stochastic3 and Stochastic5 suffer negative returns and large drawdowns. Behavioral factors also vary widely—Behavior2 and Behavior4 succeed, but Behavior3 and Behavior5 underperform with negative returns and higher drawdowns, reflecting the difficulty in consistently capturing behavioral patterns.

Although the results of factors generated from complex prompts exhibit significant variability, their overall performance is superior to those generated from simplified prompts. Complex prompts allow GPT to generate more sophisticated and targeted factors, especially in categories like trend-based analysis and mathematically complex models, where factors like Statistical1 and Statistical5 demonstrate strong annualized returns and robust Sharpe ratios. While some factors show inconsistencies or outlier behavior, particularly in trend and behavioral finance categories, the use of complex prompts generally leads to more effective and reliable factors, making them more suitable for practical financial applications.

\section{Conclusions}
In this paper, we explored the innovative application of the Generative Pre-trained Transformer (GPT) in developing factor models specifically tailored to the Chinese futures market. Our comparative analysis between GPT-generated factors and those produced by the IPCA benchmark yielded several significant findings.

Firstly, GPT-generated factors demonstrated substantial annualized returns and impressive Sharpe ratios, outperforming traditional IPCA models and showcasing superior adaptability to the dynamic financial landscape of the Chinese futures market. This highlights GPT's effectiveness in generating high-return, low-risk factors, reinforcing its potential as a powerful tool in financial modeling. Secondly, our research emphasizes the innovative potential of integrating advanced AI techniques, such as GPT, into financial modeling. By leveraging GPT’s ability to analyze and generate insights from complex datasets, we developed novel factors that effectively captured intricate market dynamics, thereby improving predictive accuracy and enhancing the effectiveness of investment strategies. Thirdly, the empirical analysis demonstrated the versatility and reliability of GPT-generated factors across various portfolio configurations, including both single and multi-factor investment strategies. This consistent performance underscores the valuable role these factors can play in modern financial strategies, offering actionable insights for financial practitioners. Finally, robustness tests confirmed that GPT-generated factors remain reliable under diverse conditions. Comparisons with other LLMs further validated GPT’s superiority in generating effective factors. Despite the inherent randomness in GPT's outputs, the factors consistently performed well. Most importantly, during the out-of-sample investment period, these factors exhibited stable and robust results, further confirming their adaptability and reliability in practical financial applications.

Our findings further validate the effectiveness of GPT in the context of futures trading, demonstrating its ability to generate reliable and high-performing factors in this distinct market segment. Building on these results, future research could extend the application of GPT-based factor modeling to other capital markets, such as options and bond markets. These markets, with their unique characteristics and data structures, present new opportunities to explore the adaptability and scalability of GPT in uncovering intricate market dynamics and generating innovative investment strategies. Such research would not only expand the scope of GPT’s applications but also provide deeper insights into its potential across diverse financial instruments. Future research could build on this foundation by examining the scalability of GPT-based factor generation across different market conditions and financial instruments, thereby broadening our understanding of its potential and limitations in global financial contexts.

\bibliographystyle{elsarticle-harv}
\bibliography{main} 
\end{document}


\maketitle

This online appendix provides supplementary materials that support the analysis in the main paper, \textit{Large Language Models and Futures Price Factors in China}. First, in Appendix A, we summarize the details of different kinds of futures contracts, including their leading and secondary codes as well as their listing dates. Second, in Appendix B, we present detailed compositions of all 40 factors generated by GPT, along with comprehensive explanations for each factor's construction. Each entry highlights the key metrics and methodologies employed, providing insights into how GPT combines various elements. Third, in Appendix C, we include empirical results from robustness checks to ensure the reliability and validity of our findings. In addition to the tables and figures referenced in these appendices, this document also contains supplementary visualizations and additional analyses that are referred to in the main paper.

\section*{Appendix A. Details of Different Futures}

For the empirical analysis conducted in our study, we collected daily market data for 104 leading and secondary continuous contracts, encompassing all stock index futures and liquid commodity futures traded in China. Elementary information for all futures contracts, including their leading and secondary codes along with listing dates, is detailed in Table \ref{DDF}.

\renewcommand{\thetable}{A\arabic{table}}
\begin{longtable}{l@{\hskip 40pt}l@{\hskip 60pt}l@{\hskip 60pt}l}
\caption{Details of Different Futures} \\ %
\label{DDF} \\ \hline
\textbf{Futures} & \textbf{Code$_{Leading}$} & \textbf{Code$_{Secondary}$} & \textbf{Listing Date} \\ \hline
\hline
\endfirsthead
\hline
\textbf{Futures} & \textbf{Code$_{Leading}$} & \textbf{Code$_{Secondary}$} & \textbf{Listing Date} \\
\hline
\endhead
\hline
\endfoot
\endlastfoot
CSI 500 & ICZL.CFE & ICZM.CFE & 2015/4/16 \\ 
CSI 300 & IFZL.CFE & IFZM.CFE & 2010/4/16 \\ 
SSE 50 & IHZL.CFE & IHZM.CFE & 2015/4/16 \\ 
CSI 1000 & IMZL.CFE & IMZM.CFE & 2022/7/22 \\ 
Apple & APZL.CZC & APZM.CZC & 2017/12/22 \\ 
Cotton & CFZL.CZC & CFZM.CZC & 2004/6/1 \\ 
Red Date & CJZL.CZC & CJZM.CZC & 2019/4/30 \\ 
Cotton Yarn & CYZL.CZC & CYZM.CZC & 2017/8/18 \\ 
Glass & FGZL.CZC & MAZM.CZC & 2012/12/3 \\ 
Methanol & MAZL.CZC & FGZM.CZC & 2014/6/17 \\ 
Vegetable Oil & OIZL.CZC & OIZM.CZC & 2012/7/16 \\ 
Staple Fiber & PFZL.CZC & PFZM.CZC & 2020/10/12 \\ 
Peanut & PKZL.CZC & PKZM.CZC & 2021/2/1 \\ 
Rapeseed Meal & RMZL.CZC & RMZM.CZC & 2012/12/28 \\ 
White Sugar & SRZL.CZC & SRZM.CZC & 2006/1/6 \\ 
PTA & TAZL.CZC & TAZM.CZC & 2006/12/18 \\ 
Urea & URZL.CZC & URZM.CZC & 2019/8/9 \\ 
Bean One & AZL.DCE & AZM.DCE & 1998/7/15 \\ 
Bean Two & BZL.DCE & BZM.DCE & 2004/1/16 \\ 
Cornstarch & CSZL.DCE & CSZM.DCE & 2014/12/19 \\ 
Corn & CZL.DCE & CZM.DCE & 2004/9/22 \\ 
Styrene & EBZL.DCE & EBZM.DCE & 2019/9/26 \\ 
Ethylene Glycol & EGZL.DCE & EGZM.DCE & 2018/12/10 \\ 
Iron Ore & IZL.DCE & IZM.DCE & 2013/10/18 \\ 
Egg & JDZL.DCE & JDZM.DCE & 2013/11/8 \\ 
Coking Coal & JMZL.DCE & JMZM.DCE & 2013/3/22 \\ 
Coke & JZL.DCE & JZM.DCE & 2011/4/15 \\ 
Pig & LHZL.DCE & LHZM.DCE & 2021/1/8 \\ 
Plastic & LZL.DCE & LZM.DCE & 2007/7/31 \\ 
Soybean Meal & MZL.DCE & MZM.DCE & 2000/7/17 \\ 
LPG & PGZL.DCE & PGZM.DCE & 2020/3/30 \\ 
Polypropylene & PPZL.DCE & PPZM.DCE & 2014/2/28 \\ 
Palm Oil & PZL.DCE & PZM.DCE & 2007/10/29 \\ 
PVC & VZL.DCE & VZM.DCE & 2009/5/25 \\ 
Soybean Oil & YZL.DCE & YZM.DCE & 2006/1/9 \\ 
Asphalt & BUZL.SHF & BUZM.SHF & 2013/10/9 \\ 
Hot Roll & HCZL.SHF & HCZM.SHF & 2014/3/21 \\ 
Low-sulfur Fuel & LUZL.SHF & LUZM.SHF & 2020/6/22 \\ 
No. 20 Rubber & NRZL.SHF & NRZM.SHF & 2019/8/12 \\ 
Rebar & RBZL.SHF & RBZM.SHF & 2009/3/27 \\ 
Rubber & RUZL.SHF & RUZM.SHF & 1997/1/16 \\ 
Crude oil & SCZL.SHF & SCZM.SHF & 2018/3/26 \\ 
Pulp & SPZL.SHF & SPZM.SHF & 2018/11/27 \\ 
Stainless Steel & SSZL.SHF & SSZM.SHF & 2019/9/25 \\ 
Shanghai Sliver & AGZL.SHF & AGZM.SHF & 2012/5/10 \\ 
Shanghai Aluminum & ALZL.SHF & ALZM.SHF & 1997/1/16 \\ 
Shanghai Gold & AUZL.SHF & AUZM.SHF & 2008/1/9 \\ 
Shanghai Copper & CUZL.SHF & CUZM.SHF & 1997/1/16 \\ 
Shanghai Nickel & NIZL.SHF & NIZM.SHF & 2015/3/27 \\ 
Shanghai Lead & PBZL.SHF & PBZM.SHF & 2011/3/24 \\ 
Shanghai Tin & SNZL.SHF & SNZM.SHF & 2015/3/27 \\ 
Shanghai Zinc & ZNZL.SHF & ZNZM.SHF & 2007/3/26 \\
\hline
\multicolumn{4}{p{\textwidth}}{Notes: This table summarizes the details of different kinds of futures contracts, including their leading and secondary codes as well as their listing dates.} \\
\end{longtable}

\section*{Appendix B. Details of Different Factors}

We provide a detailed description of each factor in Table \ref{DDFF}, categorizing them by their functional components—momentum, volatility, sentiment, liquidity, and others. For each factor listed, the appendix describes the computational methods and the data sources used, providing a thorough understanding of the mechanics underlying each indicator. By breaking down the calculation methodologies and the rationale behind each component, the factors reveal diverse aspects of market behavior. 

\renewcommand{\thetable}{B\arabic{table}}
\setcounter{table}{0}

\begin{longtable}[c]{p{5cm} p{11cm} }
    \captionsetup{justification=raggedright,singlelinecheck=false}
    \caption{Detailed Descriptions of the Financial Factors Developed and Analyzed} \\
    \label{DDFF} \\ \hline
    \textbf{Factor} & \textbf{Description} \\ \hline
    \hline
    \endfirsthead
    \multicolumn{2}{c}\
    {{\bfseries \tablename\ \thetable{} - Continued.}} \\
    \hline
    \textbf{Factor} & \textbf{Description} \\
    \hline
    \endhead
    \hline \multicolumn{2}{r}{{\textit{(Continued).}}} \\
    \endfoot
    \endlastfoot
    Innovative Momentum Volatility Sentiment Indicator (IMVSI) & Calculate momentum using the difference between the 5-day and 20-day moving averages of closing prices. Determine volatility with the 10-day standard deviation of daily returns. Normalize these measures and market sentiment, based on the futures premium to spot price ratio, and combine them to form the IMVSI factor. \\
    Adjusted Liquidity Oscillation Weighted Sentiment (ALOWS) & Calculate liquidity as volume divided by the opening price, price oscillation from the 14-day ATR divided by the spot price, and adjusted sentiment from the 10-day EMA of the futures premium and discount ratio. Combine and normalize these metrics to form the ALOWS factor. \\
    Comprehensive Market Dynamics Indicator (CMDI) & Calculate price trend strength from the absolute difference between the closing price and the 60-day moving average. Determine volume volatility and market volatility using the 30-day rolling standard deviation of volume and percentage price changes, respectively. Multiply these components to form the CMDI factor. \\
    Futures Sentiment Indicator (FSI) & Calculate Basis Impact (BI), futures Premium/Discount Impact (FPDI), Price Momentum (PM), and Volume and Amount Impact (VAI). Combine and normalize these metrics to create the FSI as a market sentiment indicator. \\
    Autocorrelation Anomaly Trend Indicator (AATI) & Calculate the Anomaly Volatility Indicator as the ratio of 10-day to 30-day rolling standard deviations, and the Trend Continuity Indicator as the ratio of short-term to long-term price changes. Combine these indicators by multiplying them to form the AATI factor. \\
    Market Reversal and Sentiment Indicator (MRSI) & Calculate extreme price movement as the sum of deviations from the 10-day maximum and minimum prices, and volume anomaly as the deviation from the 10-day average volume. Combine these with the sentiment score derived from the correlation of price and volume changes by multiplying them to form the MRSI factor. \\
    Adjusted Basis Momentum Indicator (ABMI) & Calculate Basis Movement (BM) as the difference from the previous day's basis, and Volume-Adjusted Basis Change (VABC) by normalizing BM with volume. Combine BM, VABC, and Price Momentum (PM) into the ABMI, normalize it for better interpretation, and use it as an indicator of adjusted basis momentum in the futures market.\\
    Futures Positioning Signal (FPS) & Calculate the Basis Strength Index (BSI) as the basis divided by the spot price and apply a 7-day rolling mean, and Volume Intensity (VI) as the ratio of volume to its 7-day average. Compute Price Volatility (PV) as the daily range divided by the average of the open and close prices, normalize these components, and combine them into the FPS, which is further normalized for interpretation.\\
    Trend and Volume Analysis (FTVA) & Calculate the Trend Score (TS) as the position of the close within the daily range, applying a 7-day rolling mean, and Volume Confirmation (VC) as the ratio of volume to its 7-day average, considering only increases. Combine these components, normalize them, and use them to form the FTVA, which provides an indicator of market trend and volume conviction.\\
    Market Momentum and Liquidity Indicator (MMLI) & Calculate the Momentum Score (MS) as the 7-day rolling mean of the daily price movement percentage, and the Liquidity Ratio (LR) as volume divided by amount, normalized. Combine these normalized scores and further normalize the result to form the MMLI, providing an indicator of market momentum and liquidity.\\
    Futures Price Dynamics Indicator (FPDI) & Calculate the Basis Movement Index (BMI) as the change in normalized basis (basis/spot), and the Price Range Volume Index (PRVI) as the product of normalized daily range (high-low/close) and volume, normalized. Combine BMI and PRVI, normalize the result, and form the FPDI to provide a standardized measure of futures market price dynamics.\\
    Futures Market Sentiment Indicator (FMSI) & Calculate the Premium/Discount Factor (PDF) as the percentage of the futures premium or discount relative to the spot price, and the Volume-Weighted Price Movement (VWPM) as the sum of volume-weighted daily price movements over 7 days, normalized by total volume. Combine PDF and normalized VWPM, normalize the result, and form the FMSI to provide a standardized indicator of market sentiment.\\
    Liquidity and Price Action Indicator (LPAI) & Calculate Intraday Price Amplitude as the ratio of the high-low range to the open price, Closing Price Pressure Index as the deviation of the close price from the mean of the high and low prices, Volume-Weighted Deviation as the deviation of the close price from the 10-day average weighted by volume, and Amount Momentum Trend as the percentage change in trading amount. Combine these components by multiplying them to form the LPAI.\\
    Futures Price Momentum Indicator (FPMI) & Calculate Price Momentum (PM) as the 5-day percentage change in the closing price, and Volume Momentum (VM) as the 5-day percentage change in volume. Combine these changes to form the FPMI, and optionally normalize it to a z-score for better comparison over time.\\
    Futures Basis Volatility Indicator (FBVI) &	Calculate Basis Volatility (BV) as the 10-day rolling standard deviation of the basis, and normalize volume to obtain a Volume Weight (VW). Combine BV and VW equally, normalize the result, and form the FBVI to compare across different futures contracts or timeframes.\\
    Fundamental Value and Market Distortion Indicator (FVMDI) & Calculate Basis Value Deviation as the deviation of basis from spot price, futures Premium/Discount Adjustment as the absolute value of the futures premium and discount, and Price Stability Index as the inverse of the daily price range. Combine these with Trading Momentum Amount Correlation, which measures the correlation between volume momentum and amount change, and multiply them to form the FVMDI.\\
    Stability and Volume Indicator (SVI) & Calculate Price Stability as the inverse of the 20-day rolling standard deviation of price changes, and Volume Change as the 10-day percentage change in volume. Combine these with the 20-day rolling correlation between price and volume changes to form the SVI, representing market stability and volume dynamics.\\
    Volatility Asymmetry Synergy Indicator (VASI) & Calculate the Volatility Ratio as the ratio of short-term (5-day) to long-term (30-day) volatility, and Price Asymmetry as the ratio of the sum of positive price changes to the sum of negative price changes over 5 days. Combine these with Volume Synergy, the 5-day rolling correlation between price and volume changes, to form the VASI.\\
    Futures Market Sentiment Oscillator (FMSO) & Calculate the Spot-Premium Ratio (SPR) as the spot price divided by the sum of the spot price and futures premium/discount, and the Volume Impact Factor (VIF) by normalizing volume by its maximum. Combine SPR and VIF to form the FMSO, and normalize it to get the FMSO Index, representing the market sentiment score.\\
    Futures Insight Gauge (FIG) & Calculate the Daily Price Range (DPR) as the ratio of the high-low range to the open price, and the Premium/Discount Influence (PDI) as the futures premium/discount divided by the spot price. Combine DPR and PDI with the Volume Significance Score (VSS), normalized based on maximum volume and DPR, to form the FIG, which is further normalized to create the FIG Index, representing the market insight score.\\
    Liquidity Shock Indicator (LSI) & Calculate the Daily Price Range (high minus low) and the Daily Volume Change Percentage for each futures contract. Smooth both using a rolling standard deviation and normalize them with z-scores. The Liquidity Shock Indicator (LSI) is the product of these z-scores, highlighting periods of high volatility and volume shifts. Finally, normalize the LSI to create the LSI Factor, representing the liquidity shock signal for each contract.\\
    Market Momentum Meter (MMM) & Calculate the Premium/Discount Momentum (PDM) as the rate of change of the futures premium/discount, the Volatility Gauge (VG) as the high-low range, and the Volume Trend Indicator (VTI) as the change in volume. Normalize these components, combine them equally to form the MMM, providing a momentum score for each futures contract.\\
    Price Stability and Volume Coordination Indicator (PSVCI) & To compute the Price Stability and Volume Coordination Indicator (PSVCI), the functions for intraday price stability, volume change rate, and price-volume coordination are applied to each group of data based on ‘futuresname’. The final PSVCI is the product of these three computed values.\\
    Basis Volatility Impact Factor (BVI) & To compute the Basis Volatility Impact Factor (BVI), calculate the standardized basis volatility, the price-basis correlation, and the volume adjustment. The BVI is then obtained by multiplying these three components together.\\
    Futures Market Analytics Tool (FMAT) & To compute the futures Market Analytics Tool (FMAT), calculate the Daily Price Movement Indicator (DPMI), the Basis Indicator (BI), and the Liquidity Indicator (LI). Normalize these indicators, then average the normalized values to create the FMAT index.\\
    Basis Volatility Spread (BVS) & To compute the Basis Volatility Spread (BVS), calculate the Basis Volatility (BV) using a rolling standard deviation of the basis and the Mean futures Premium/Discount (MFPD) using a rolling mean. The BVS is then obtained by subtracting the MFPD from the BV.\\
    Multidimensional Market Dynamics Indicator (MMDI) & To compute the Multidimensional Market Dynamics Indicator (MMDI), calculate the price momentum, volume dynamics, and adjusted volatility for each group of data based on ‘futuresname’. The final MMDI is obtained by multiplying these three components together.\\
    Momentum-Volume Oscillator (MVO) & To compute the Momentum-Volume Oscillator (MVO), calculate the price momentum, volume dynamics, and volatility for each group of data based on ‘futuresname’. The MVO is then obtained by multiplying these three components together.\\
    Basis-Adjusted Volume Oscillator (BAVO) & To compute the Basis-Adjusted Volume Oscillator (BAVO), calculate the short-term and long-term exponential moving averages for volume, then compute the volume oscillator as their difference. Normalize the basis by dividing it by the spot price. Finally, compute BAVO by multiplying the volume oscillator with the normalized basis.\\
    Spot-Relative Momentum Indicator (SRMI) & To compute the Spot-Relative Momentum Indicator (SRMI), first calculate the daily percentage changes for futures and spot prices. Then, compute the daily price change ratio and adjust it by the volume-weighted ratio. Finally, apply an exponential moving average to the volume-weighted ratio to obtain the SRMI.\\
    Behavioral Momentum Indicator (BMI) & To compute the Behavioral Momentum Indicator (BMI), calculate the trend intensity, volume acceleration, and price acceleration for each group based on ‘futuresname’. The BMI is then obtained by multiplying the trend intensity with the sum of volume acceleration and price acceleration.\\
    Basis Momentum Volatility Indicator (BMVI) & To compute the Basis Momentum Volatility Indicator (BMVI), first calculate the basis volatility, momentum score, and average volume over a 20-day period. Then, compute the volume amplification by dividing the current volume by the average volume. Finally, obtain the BMVI by multiplying the basis volatility, momentum score, and volume amplification.\\
    Spot-futures Convergence Divergence (SFCD) & To compute the Spot-futures Convergence Divergence (SFCD), first calculate the moving average of the basis and the Convergence Divergence Ratio (CDR) by dividing the basis by its moving average. Next, compute the volume trend as the percentage change in volume over 10 periods. Finally, obtain the SFCD by multiplying the CDR by the volume trend.\\
    Price Range Trend Indicator (FPRTI) & To compute the futures Price Range Trend Indicator (FPRTI), first calculate the Daily Price Range (DPR) and the Price Range Movement (PRM). Then, compute the average volume over a specified period and determine the volume weighting. Finally, obtain the FPRTI by multiplying the DPR, PRM, and volume weighting.\\
    Advanced Market Efficiency Indicator (AMEI) & To compute the Advanced Market Efficiency Indicator (AMEI), calculate the efficiency ratio, normalized deviation from the moving average, and volatility adjustment for each group based on ‘futuresname’. The AMEI is then obtained by multiplying the efficiency ratio and the normalized deviation from the moving average, divided by the volatility adjustment.\\
    Futures Momentum Score (FMS) & For the futures Momentum Score (FMS), calculate the price momentum, volume intensity, and volatility range. Normalize these components, then combine them by averaging to create the FMS.\\
    Dynamic Market Pulse (DMP) & For the Dynamic Market Pulse (DMP), calculate the high-low average, volume average, STVT, and LSI. The DMP is obtained by multiplying STVT and LSI, then dividing by the basis.\\
    Futures Close-to-Spot Ratio (FCSR) & To compute the futures Close-to-Spot Ratio (FCSR), calculate the ratio of the closing price to the spot price and subtract 1.\\
    Composite Market Strength Indicator (CMSI) & For the Composite Market Strength Indicator (CMSI), calculate the price momentum, Relative Strength Index (RSI), and sentiment score. Combine these components using specified weights to obtain the CMSI.\\
    Futures Liquidity-Adjusted Basis Trend (FLABT) & To compute the futures Liquidity-Adjusted Basis Trend (FLABT), calculate the exponential moving average of the basis, determine the liquidity ratio by normalizing volume and amount, and compute the adjusted basis trend. Finally, normalize the adjusted basis trend to obtain the FLABT score.\\
    \hline
    \multicolumn{2}{p{\dimexpr\textwidth-2\tabcolsep\relax}}{Notes: This table presents the detailed compositions of all 40 factors generated by GPT, along with comprehensive explanations for each factor's construction. Each entry highlights the key metrics and methodologies employed, providing insights into how GPT combines various elements.} \\
\end{longtable}

\section*{Appendix C. Robustness}
This section presents the results of all robustness checks performed, including comparisons of outcomes from different LLMs, evaluations of various dialogue window settings, and analyses of different prompt engineering strategies. The tables in this appendix provide comprehensive empirical results that underscore the reliability and consistency of our findings under diverse conditions.

\subsection*{C.1 Different LLMs Comparison}

Table \ref{PC} in this subsection provides a detailed comparison of the performance across different LLMs used in our study, including GPT-4o, Copilot, and Claude 3.5. 

\renewcommand{\thetable}{C\arabic{table}}
\setcounter{table}{0}

\begin{table}[H]
\centering
\begin{threeparttable}
\caption{Performance Comparison of Different LLMs}
\label{PC}
\vspace{1em} 
{\footnotesize
\begin{tabularx}{\textwidth}{l *{3}{>{\centering\arraybackslash}X}}
\toprule
\multicolumn{4}{c}{\textbf{Long-short Portfolio}} \\
\midrule
\textbf{Factor} & \textbf{Annualized return} & \textbf{Sharpe ratio} & \textbf{Max drawdown} \\
\midrule
GPT4o-1 & -0.1887 & -0.6528 & 0.5918 \\
GPT4o-2 & 6.9277 & 11.6070 & 0.0461 \\
GPT4o-3 & 3.3906 & 7.3329 & 0.0675 \\
GPT4o-4 & -0.1108 & -0.6416 & 0.1801 \\
GPT4o-5 & -0.4469 & -2.6526 & 0.5365 \\
Copilot-1 & -0.0455 & -0.0877 & 0.3171 \\
Copilot-2 & -0.1830 & -0.8665 & 0.3649 \\
Copilot-3 & -0.1280 & -0.6412 & 0.2767 \\
Copilot-4 & -0.9616 & -15.5601 & 0.9804 \\
Copilot-5 & -0.0470 & -0.1225 & 0.3219 \\
Claude-1 & -0.1292 & -0.4403 & 0.3338 \\
Claude-2 & -0.3087 & -2.6150 & 0.5070 \\
Claude-3 & 0.3916 & 1.3824 & 0.1781 \\
Claude-4 & 0.0639 & 0.5401 & 0.1482 \\
Claude-5 & 1.1060 & 4.9952 & 0.0752 \\
\midrule
\multicolumn{4}{>{\hsize=\textwidth\centering}X}{\textbf{Long-Only Portfolio}} \\
\midrule
\textbf{Factor} & \textbf{Annualized return} & \textbf{Sharpe ratio} & \textbf{Max drawdown} \\
\midrule
GPT4o-1 & -0.1497 & -0.3661 & 0.5307 \\
GPT4o-2 & 1.7446 & 5.5843 & 0.0750 \\
GPT4o-3 & 1.0688 & 3.7907 & 0.0908 \\
GPT4o-4 & 0.4726 & 1.9032 & 0.1371 \\
GPT4o-5 & -0.2754 & -1.2649 & 0.4291 \\
Copilot-1 & -0.0607 & -0.0745 & 0.4019 \\
Copilot-2 & -0.0316 & -0.0613 & 0.3173 \\
Copilot-3 & -0.0943 & -0.2047 & 0.3797 \\
Copilot-4 & -0.8164 & -8.0018 & 0.8706 \\
Copilot-5 & -0.0259 & -0.0544 & 0.2651 \\
Claude-1 & -0.1586 & -0.4268 & 0.3939 \\
Claude-2 & -0.0693 & -0.1958 & 0.3453 \\
Claude-3 & 0.1776 & 0.8106 & 0.1521\\
Claude-4 & 0.0369 & 0.3676 & 0.1447 \\
Claude-5 & 0.3666 & 2.3108 & 0.0843 \\
\bottomrule
\end{tabularx}
}
\begin{tablenotes}[flushleft]
\item Notes: The table displays the returns of factors generated by GPT in different LLMs when implemented in long-short and long-only portfolios. 
\end{tablenotes}
\end{threeparttable}
\end{table}

\subsection*{C.2 Parameter Temperature Variations}

Table \ref{PNCLS} in this subsection presents the impact of parameter temperature variations on the performance of financial factors generated by LLMs. We conducted an experiment across ten independent dialogue conversations, each generating five factors, to assess the robustness and consistency of the LLM-generated factors under different temperature settings. 

\begin{table}[H]
\centering
\caption{Performance of Factors From New Conversation Windows: Long-Short Portfolio}
\label{PNCLS}
\vspace{1em} 
\resizebox{\textwidth}{!}{
\begin{threeparttable}
\begin{tabular}{l c c c l c c c}
\hline
\textbf{Factor} & \textbf{Annualized} & \textbf{Sharpe} & \textbf{Max} & \textbf{Factor} & \textbf{Annualized} & \textbf{Sharpe} & \textbf{Max} \\
& \textbf{Return} & \textbf{Ratio} & \textbf{Drawdown} & & \textbf{Return} & \textbf{Ratio} & \textbf{Drawdown} \\
\hline
Factor1-1 & 2.4588 & 5.6106 & 0.0853 & Factor6-1 & 13.2776 & 13.8652 & 0.0330 \\
Factor1-2 & 24.3286 & 16.4453 & 0.0436 & Factor6-2 & 1.2465 & 4.2835 & 0.0936 \\
Factor1-3 & 0.0110 & 0.1593 & 0.2337 & Factor6-3 & -0.0882 & -0.3455 & 0.2031 \\
Factor1-4 & -0.0224 & -0.0306 & 0.1960 & Factor6-4 & -0.4757 & -2.0454 & 0.5878 \\
Factor1-5 & 0.3828 & 1.6868 & 0.1323 & Factor6-5 & 0.0965 & 0.7253 & 0.1477 \\
Factor2-1 & -0.0120 & 0.0976 & 0.3467 & Factor7-1 & 12.2613 & 11.8802 & 0.0526 \\
Factor2-2 & -0.5450 & -3.9831 & 0.6688 & Factor7-2 & -0.3368 & -1.1400 & 0.5056 \\
Factor2-3 & 8.1339 & 8.0091 & 0.0906 & Factor7-3 & -0.1455 & -0.5741 & 0.3830 \\
Factor2-4 & -0.0456 & -0.1846 & 0.1889 & Factor7-4 & 0.0712 & 0.4785 & 0.1406 \\
Factor2-5 & 0.0669 & 0.4293 & 0.2045 & Factor7-5 & -0.1233 & -0.7781 & 0.2102 \\
Factor3-1 & 0.0286 & 0.2374 & 0.2268 & Factor8-1 & -0.1239 & -0.6910 & 0.2685 \\
Factor3-2 & -0.1397 & -0.9066 & 0.2079 & Factor8-2 & 1.4917 & 5.1852 & 0.0547 \\
Factor3-3 & -0.5642 & -4.8172 & 0.6476 & Factor8-3 & 7.0147 & 8.6627 & 0.0961 \\
Factor3-4 & 14.0715 & 12.6639 & 0.0486 & Factor8-4 & -0.6836 & -6.1337 & 0.7505 \\
Factor3-5 & -0.1927 & -0.5441 & 0.4617 & Factor8-5 & -0.1704 & -0.7330 & 0.2912 \\
Factor4-1 & -0.1417 & -0.5310 & 0.3028 & Factor9-1 & 9.1250 & 12.8986 & 0.0581 \\
Factor4-2 & 1.6141 & 4.2909 & 0.1719 & Factor9-2 & 0.0826 & 0.4654 & 0.3494 \\
Factor4-3 & 0.0096 & 0.1381 & 0.1339 & Factor9-3 & 0.0940 & 0.8890 & 0.0869 \\
Factor4-4 & -0.3839 & -1.9608 & 0.5221 & Factor9-4 & -0.2151 & -1.0789 & 0.2972 \\
Factor4-5 & 13.5866 & 13.4466 & 0.0354 & Factor9-5 & -0.5269 & -3.7816 & 0.6061 \\
Factor5-1 & 15.0598 & 12.8800 & 0.0448 & Factor10-1 & -0.0962 & -0.5373 & 0.2141 \\
Factor5-2 & 15.0598 & 12.8800 & 0.0448 & Factor10-2 & -0.0962 & -0.5373 & 0.2141 \\
Factor5-3 & -0.0449 & -0.1590 & 0.2626 & Factor10-3 & 0.0872 & 0.6041 & 0.1257 \\
Factor5-4 & -0.0839 & -0.5491 & 0.2079 & Factor10-4 & -0.0962 & -0.5373 & 0.2141 \\
Factor5-5 & -0.0215 & 0.0433 & 0.5725 & Factor10-5 & -0.0616 & -0.2428 & 0.1825 \\
\hline
\end{tabular}
\begin{tablenotes}[flushleft]
\item Notes: The table displays the returns of factors generated by GPT from new conversation windows when implemented in long-short portfolios. 
\end{tablenotes}
\end{threeparttable}
}
\end{table}

\begin{table}[H]
\centering
\caption{Performance of Factors From New Conversation Windows: Long-Only Portfolio}
\label{PNCLO}
\vspace{1em} 
\resizebox{\textwidth}{!}{
\begin{threeparttable}
\begin{tabular}{l c c c l c c c}
\hline
\textbf{Factor} & \textbf{Annualized} & \textbf{Sharpe} & \textbf{Max} & \textbf{Factor} & \textbf{Annualized} & \textbf{Sharpe} & \textbf{Max} \\
& \textbf{Return} & \textbf{Ratio} & \textbf{Drawdown} & & \textbf{Return} & \textbf{Ratio} & \textbf{Drawdown} \\
\hline
Factor1-1 & 0.9526 & 3.6155 & 0.0978 & Factor6-1 & 2.4655 & 7.6252 & 0.0486 \\
Factor1-2 & 3.9703 & 7.5723 & 0.0681 & Factor6-2 & 0.5573 & 3.5662 & 0.0689 \\
Factor1-3 & -0.0525 & -0.0689 & 0.3035 & Factor6-3 & 0.0176 & 0.1916 & 0.2983 \\
Factor1-4 & -0.0990 & -0.7361 & 0.2158 & Factor6-4 & -0.2619 & -0.6992 & 0.4818 \\
Factor1-5 & 0.0903 & 0.5945 & 0.1547 & Factor6-5 & -0.0261 & -0.0142 & 0.3000 \\
Factor2-1 & -0.0856 & -0.1127 & 0.3857 & Factor7-1 & 2.5459 & 6.8111 & 0.0711 \\
Factor2-2 & -0.3736 & -1.9422 & 0.5243 & Factor7-2 & -0.3571 & -1.0538 & 0.5489 \\
Factor2-3 & 2.0465 & 4.6746 & 0.1093 & Factor7-3 & -0.0493 & -0.1763 & 0.2801 \\
Factor2-4 & -0.0443 & -0.2014 & 0.2302 & Factor7-4 & -0.0235 & -0.0570 & 0.1978 \\
Factor2-5 & 0.0463 & 0.3113 & 0.2705 & Factor7-5 & -0.1224 & -0.9986 & 0.2114 \\
Factor3-1 & -0.0344 & 0.0239 & 0.3451 & Factor8-1 & 0.0457 & 0.3444 & 0.2332 \\
Factor3-2 & -0.0732 & -0.3743 & 0.2365 & Factor8-2 & 0.6550 & 3.8222 & 0.0488 \\
Factor3-3 & -0.4775 & -2.8269 & 0.6253 & Factor8-3 & 1.8186 & 4.8957 & 0.1171 \\
Factor3-4 & 2.7586 & 5.7915 & 0.1233 & Factor8-4 & -0.4749 & -3.1249 & 0.5402 \\
Factor3-5 & -0.2237 & -0.6340 & 0.5165 & Factor8-5 & -0.0715 & -0.1398 & 0.3293 \\
Factor4-1 & -0.0583 & -0.1743 & 0.2992 & Factor9-1 & 2.1874 & 6.0373 & 0.0849 \\
Factor4-2 & 0.5807 & 2.2903 & 0.1398 & Factor9-2 & 0.0372 & 0.2897 & 0.1523 \\
Factor4-3 & -0.0346 & -0.0964 & 0.2236 & Factor9-3 & -0.0437 & -0.3156 & 0.1862 \\
Factor4-4 & -0.2695 & -0.9904 & 0.4393 & Factor9-4 & -0.0670 & -0.5134 & 0.1656 \\
Factor4-5 & 2.0764 & 4.9754 & 0.1231 & Factor9-5 & -0.1685 & -0.6958 & 0.3618 \\
Factor5-1 & 2.6920 & 7.3562 & 0.0753 & Factor10-1 & -0.1143 & -0.7847 & 0.2425 \\
Factor5-2 & 2.6920 & 7.3562 & 0.0753 & Factor10-2 & -0.1143 & -0.7847 & 0.2425 \\
Factor5-3 & -0.0959 & -0.5220 & 0.3171 & Factor10-3 & 0.0570 & 0.3729 & 0.2327 \\
Factor5-4 & 0.0433 & 0.3166 & 0.2173 & Factor10-4 & -0.1143 & -0.7847 & 0.2425 \\
Factor5-5 & 0.0734 & 0.3937 & 0.4996 & Factor10-5 & -0.1106 & -0.4644 & 0.2458 \\
\hline
\end{tabular}
\begin{tablenotes}[flushleft]
\item Notes: The table displays the returns of factors generated by GPT from new conversation windows when implemented in long-only portfolios. 
\end{tablenotes}
\end{threeparttable}
}
\end{table}

\subsection*{C.3 Prompt Engineering}
Table \ref{PGS} and Table \ref{PGC} in this subsection present the impact of prompt engineering on the performance of LLMs in generating financial factors. We divide this analysis into two parts: simple prompts and specific prompts. Each part evaluates how different levels of prompt complexity influence the robustness, stability, and overall performance of the generated factors.

\begin{table}[htbp]
\centering
\begin{threeparttable}
\caption{Performance of Factors Given Simple Prompts}
\label{PGS}
\vspace{1em} 
\begin{tabularx}{\textwidth}{l *{3}{>{\centering\arraybackslash}X}}
\toprule
\multicolumn{4}{c}{\textbf{Long-short Portfolio}} \\
\midrule
\textbf{Factor} & \textbf{Annualized return} & \textbf{Sharpe ratio} & \textbf{Max drawdown} \\
\midrule
Simple1 & -0.1323 & -0.5878 & 0.2100 \\
Simple2 & -0.1227 & -0.7811 & 0.2324 \\
Simple3 & -0.1118 & -0.4083 & 0.3743 \\
Simple4 & 4.6323 & 7.9456 & 0.0860 \\
Simple5 & -0.1814 & -0.8095 & 0.3029 \\
Simple6 & 9.4187 & 9.0420 & 0.1245 \\
Simple7 & 9.3493 & 12.3554 & 0.0607 \\
Simple8 & -0.1882 & -0.8692 & 0.2661 \\
Simple9 & 0.1059 & 0.7074 & 0.1406 \\
Simple10 & -0.3975 & -1.7835 & 0.5222 \\
Simple11 & 0.0382 & 0.5119 & 0.1328 \\
Simple12 & -0.2108 & -1.0184 & 0.2696 \\
Simple13 & -0.1990 & -0.9689 & 0.2564 \\
Simple14 & -0.1880 & -0.8836 & 0.2464 \\
Simple15 & 5.3381 & 10.7164 & 0.0461 \\
\midrule
\multicolumn{4}{>{\hsize=\textwidth\centering}X}{\textbf{Long-Only Portfolio}} \\
\midrule
\textbf{Factor} & \textbf{Annualized return} & \textbf{Sharpe ratio} & \textbf{Max drawdown} \\
\midrule
Simple1 & -0.0421 & -0.0650 & 0.3145 \\
Simple2 & -0.1113 & -0.6085 & 0.2564 \\
Simple3 & -0.1096 & -0.5248 & 0.3334 \\
Simple4 & 1.3104 & 3.9366 & 0.1122 \\
Simple5 & -0.1012 & -0.2836 & 0.3442 \\
Simple6 & 2.4759 & 5.4668 & 0.1352 \\
Simple7 & 2.0213 & 6.7414 & 0.0701 \\
Simple8 & -0.0933 & -0.2414 & 0.3298 \\
Simple9 & -0.0404 & -0.1251 & 0.2365 \\
Simple10 & -0.1811 & -0.7472 & 0.3884 \\
Simple11 & 0.9329 & 0.9561 & 0.0307 \\
Simple12 & -0.0797 & -0.2285 & 0.3270 \\
Simple13 & -0.0935 & -0.2873 & 0.3333 \\
Simple14 & -0.0810 & -0.2363 & 0.3242 \\
Simple15 & 1.4684 & 5.8674 & 0.0617 \\
\bottomrule
\end{tabularx}
\begin{tablenotes}[flushleft]
\item Notes: The table displays the returns of factors generated by GPT given simple prompts when implemented in long-short and long-only portfolios. 
\end{tablenotes}
\end{threeparttable}
\end{table}

\begin{table}[htbp]
\centering
\begin{threeparttable}
\caption{Performance of Factors Given Specific Prompts}
\label{PGC}
\vspace{1em} 
{\footnotesize
\begin{tabularx}{\textwidth}{l *{3}{>{\centering\arraybackslash}X}}
\toprule
\multicolumn{4}{c}{\textbf{Long-short Portfolio}} \\
\midrule
\textbf{Factor} & \textbf{Annualized return} & \textbf{Sharpe ratio} & \textbf{Max drawdown} \\
\midrule
Trend1 & -0.0312 & -0.05417 & 0.2585 \\
Trend2 & 0.7230 & 2.2880 & 0.1372 \\
Trend3 & 0.1837 & 1.1076 & 0.2939 \\
Trend4 & 0.2202 & 0.9249 & 0.6782 \\
Trend5 & -0.0768 & -0.2870 & 0.3045 \\
Statistical1 & 13.6568 & 13.4455 & 0.0484 \\
Statistical2 & 0.0092 & 0.1424 & 0.3446 \\
Statistical3 & -0.1646 & -0.6962 & 0.3127 \\
Statistical4 & 0.0677 & 0.4414 & 0.1624 \\
Statistical5 & 33.9494 & 16.2141 & 0.0509 \\
Stochastic1 & 0.0129 & 0.1601 & 0.1575 \\
Stochastic2 & 0.0284 & 0.2842 & 0.1043 \\
Stochastic3 & -0.0635 & -0.5361 & 0.1652 \\
Stochastic4 & -0.1378 & -0.4882 & 0.3175 \\
Stochastic5 & -0.0824 & -0.4605 & 0.2045 \\
Behavior1 & -0.0492 & -0.2086 & 0.2161 \\
Behavior2 & 0.4937 & 1.6444 & 0.1892 \\
Behavior3 & -0.1951 & -0.8163 & 0.3537 \\
Behavior4 & 15.9041 & 13.7138 & 0.0481 \\
Behavior5 & -0.0620 & -0.1651 & 0.1874 \\
\midrule
\multicolumn{4}{>{\hsize=\textwidth\centering}X}{\textbf{Long-Only Portfolio}} \\
\midrule
\textbf{Factor} & \textbf{Annualized return} & \textbf{Sharpe ratio} & \textbf{Max drawdown} \\
\midrule
Trend1 & -0.0528 & -0.1664 & 0.2970 \\
Trend2 & 0.3094 & 1.3900 & 0.1324 \\
Trend3 & -0.0041 & 0.0727 & 0.3502 \\
Trend4 & 0.0625 & 0.3643 & 0.5332 \\
Trend5 & -0.0005 & 0.0951 & 0.2733 \\
Statistical1 & 2.4709 & 6.8362 & 0.0870 \\
Statistical2 & 0.0367 & 0.2698 & 0.3556 \\
Statistical3 & -0.0665 & -0.2387 & 0.2337 \\
Statistical4 & -0.0843 & -0.2793 & 0.3028 \\
Statistical5 & 4.7083 & 9.0429 & 0.0533 \\
Stochastic1 & -0.0251 & -0.0724 & 0.1682 \\
Stochastic2 & 0.0084 & 0.1367 & 0.1719 \\
Stochastic3 & -0.0724 & -0.4865 & 0.1907 \\
Stochastic4 & -0.0020 & 0.0223 & 0.1275 \\
Stochastic5 & -0.0566 & -0.1520 & 0.3037 \\
Behavior1 & 0.0212 & 0.2058 & 0.1966 \\
Behavior2 & 0.2292 & 0.9921 & 0.1995 \\
Behavior3 & -0.1048 & -0.2784 & 0.3214 \\
Behavior4 & 2.6670 & 5.8202 & 0.1060 \\
Behavior5 & -0.1050 & -0.3141 & 0.3050 \\
\bottomrule
\end{tabularx}
}
\begin{tablenotes}[flushleft]\footnotesize
\item Notes: The table displays the returns of specific types of factors generated by GPT when implemented in long-short and long-only portfolios. We specified the characteristics of the generated factors, including four categories: those related to trends, mathematically complex factors, factors associated with Fourier analysis, and factors correlated with behavioral finance.
\end{tablenotes}
\end{threeparttable}
\end{table}